\documentclass[10pt]{article}

\usepackage{changepage}

\usepackage[utf8]{inputenc}

\usepackage{textcomp,marvosym}

\usepackage{fixltx2e}

\usepackage{amsmath,amssymb}

\usepackage{cite}

\usepackage{nameref}

\usepackage{microtype}
\DisableLigatures[f]{encoding = *, family = * }

\usepackage{rotating}

\usepackage[percent]{overpic}

\makeatletter
\renewcommand{\@biblabel}[1]{\quad#1.}
\makeatother

\date{}

\begin{document}
\vspace*{0.35in}

\begin{flushleft}
{\Large
\textbf\newline{Do Brain Networks Evolve by Maximizing their Information Flow Capacity?}
}
\newline
\\
Chris G. Antonopoulos\textsuperscript{1*},
Shambhavi Srivastava\textsuperscript{1},
Sandro E. de S. Pinto\textsuperscript{2},
Murilo S. Baptista\textsuperscript{1}
\\
\bf{1} Department of Physics (ICSMB), University of Aberdeen, SUPA, Aberdeen, United Kingdom
\\
\bf{2} Departamento de F\'isica, Universidade Estadual de Ponta Grossa, Paran\'a, Brazil

%
%





* E-mail: Corresponding chris.antonopoulos@abdn.ac.uk
\end{flushleft}
\section*{Abstract}
We propose a working hypothesis supported by numerical simulations that brain networks evolve based on the principle of the maximization of their internal information flow capacity. We find that synchronous behavior and capacity of information flow of the evolved networks reproduce well the same behaviors observed in the brain dynamical networks of \textit{Caenorhabditis elegans} and humans, networks of Hindmarsh-Rose neurons with graphs given by these brain networks. We make a strong case to verify our hypothesis by showing that the neural networks with the closest graph distance to the brain networks of \textit{Caenorhabditis elegans} and humans are the Hindmarsh-Rose neural networks evolved with coupling strengths that maximize information flow capacity. Surprisingly, we find that global neural synchronization levels decrease during brain evolution, reflecting on an underlying global no Hebbian-like evolution process, which is driven by no Hebbian-like learning behaviors for some of the clusters during evolution, and Hebbian-like learning rules for clusters where neurons increase their synchronization.

\section*{Author Summary}
The study of the function of the brain is of primordial importance in neuroscience. Several brain models have been studied so far that take into account higher level functions of the external stimulus and the behavioral response attributed to ensembles of neurons of cortical areas. If the brain learns by maximizing the Mutual Information between stimuli and response, or by updating the internal model of probabilities by using Bayesian techniques, or by minimizing the free-energy, it provides little insight about the dynamical mechanisms appearing in brain networks when evolved based on such principles. In our work we propose a working hypothesis supported by numerical simulations that brain dynamical networks evolve based on the principle of the maximization of their internal information flow capacity, i.e. the upper bound for the information transferred per time unit between any two nodes. We make a strong case to verify our hypothesis by showing that the neural networks with the closest spectral graph distance to the brain networks of \textit{Caenorhabditis elegans} and humans are the Hindmarsh-Rose neural networks evolved with coupling strengths that maximize information flow capacity. We also find that synchronous behavior and capacity of information flow of the evolved neural networks reproduce well the same behaviors observed in the brain dynamical networks of \textit{Caenorhabditis elegans} and humans. Finally, we find that global neural synchronization levels decrease during brain network evolution.

\section*{Introduction}\label{section_introduction}

A plethora of phenomena in nature can be effectively described by networks. Neuroscientists have used tools for the analysis of complex networks that help realize even more deeply the functionality and structure of the brain. It was found that many aspects of brain network structures are typical of a wide range of non-neural or non-biological complex networks \cite{Meunieretal2001,Yamaguti2014}. One of the main findings in neuroscience is the modular organization of the brain, which in turn implies an inherent parallel nature of brain computations \cite{Meunieretal2001}. Modular processors have to be sufficiently isolated and dynamically differentiated to achieve independent computations, but also globally connected to be integrated in coherent functions \cite{Meunieretal2001}. It has been revealed that the cortical network is a hierarchical and clustered network with a complex connectivity \cite{Hilgetag_2004}. A possible network description for this modular organization is that brain networks may be small-world structured \cite{Heetal2007} with properties similar to many other complex networks \cite{Stam2004}. This viewpoint has been driven by the systematic finding of small-world topology in a wide range of human brain networks derived from structural \cite{Heetal2007}, functional \cite{Eguiluzetal2005}, and diffusion tensor MRI \cite{Hagmannetal2007} studies. Small-world topology has also been identified at the cellular-network scale in functional cortical neural circuits in mammals \cite{Yuetal2008} and also in the nervous system of the nematode \textit{Caenorhabditis elegans} (\textit{C.elegans}) \cite{Wattsetal1998}. Moreover, this topology seems to be relevant for the brain function because it is affected by diseases \cite{Yuetal2007}, normal ageing, and by pharmacological blockade of dopamine neurotransmission \cite{Achardetal2007}.

Synchronization is ubiquitous in nature. Insightful findings regarding synchronization in complex networks were reviewed recently in Ref. \cite{Arenasetal2008}. Recently, synchronization in complex modular or clustered networks has been investigated \cite{Gardenesetal2011,Viana_2014}. It appears as the interplay between the intrinsic dynamics associated to the nodes of the network and its graph topology and connecting functions. In this work, synchronization will be considered as a mean to quantify functional behaviors of the brain dynamical networks (BDNs) studied. By BDN we mean a network that represents the connectome equipped with neural dynamics for its nodes to account for their time evolution.

Mathematical and computational approaches have a long tradition traced back to the early mathematical theories of perception \cite{Helmholtz1913} and of current integration by a neural cell membrane \cite{Lapicqueetal1907}. Hebb's idea on assembly formation \cite{Hebb1949} inspired simulations on the largest computers available at that time (1956) \cite{Rochesteretal1956} to understand the relation between neural connectivity strength and response, i.e. behavior. It is a learning rule which proposes an explanation for the adaptation of neurons during the learning process. The simultaneous activation of pairs of neurons leads to pronounced increases in their synaptic strength. There is also the possibility of many other kinds of learning \cite{Gerstneretal2008}. In this work we find evidence of Hebbian-like and no Hebbian-like learning rules characterized by the relationship between the addition of synapses and the increase or decay in the synchronization behavior of neurons in the evolved BDNs.

Several brain models have been studied so far that do not take into account particular behaviors of isolated neurons, but higher level functions of the external stimulus and the behavioral response attributed to ensembles of neurons of cortical areas. There are two classes: Those based on collective functional dynamics of local groups of neurons, such as the Wilson-Cowan model for the cortical and thalamic nervous tissue \cite{wilson_cybernetic_1973}, and those based on the conditional probabilities (as well as information and mutual information (MI)) of stimuli and responses, prominent examples of which are the Bayesian brain hypothesis \cite{knill} and the infomax theory \cite{linsker_Computer1988}. Recently, it was shown that most of the probabilistic brain models can be unified under a single free-energy principle \cite{friston_nature2010}, the one that interprets the brain as a system that tries to minimize surprises of the sensations from the world.

If the brain learns by maximizing the MI between stimuli and response \cite{linsker_Computer1988}, or by updating the internal model of probabilities by using Bayesian techniques \cite{knill}, or by minimizing the free-energy \cite{friston_nature2010}, the surprise of the response provides little insight about the dynamical mechanisms appearing in a brain network when it evolves based on such principles. The drawback however of the probabilistic brain models is that they describe little about the underlying dynamical structure of what is really happening in the neural level and the representation of the stimuli \cite{luce_RGP2003}.

For deterministic systems with correlations, an appropriate quantity for measuring the transfer of information is the Mutual Information Rate (MIR), the MI per time unit. In Ref. \cite{Baptistaetal2012}, the authors have developed alternative methods to overcome problems that stem from the definition of probabilities from time series and derived an upper bound for the MIR, $I_c$, between two nodes of a complex dynamical network from time averages that do not rely on probabilities, but instead on the two largest Lyapunov exponents $l_1$, $l_2$ of the subspace of the network formed by the two nodes (see Eq. \eqref{Ic_MIR} in Materials and Methods). In our study, $I_c$ stands for the upper bound for the information transferred per time unit between any two nodes of the BDN, what represents the information flow capacity of the BDN. We discuss more on these in Materials and Methods, Section Upper Bound for MIR.

Inspired by the infomax theory and by theoretical studies that proposes that the maximization of information transmission between subsystems can be used as a principle for understanding the development and evolution of complex brain networks (see Ref. \cite{Yamaguti2014} and references therein, and Ref. \cite{Tsuda2015}) and, aiming at elucidating the microscopic dynamic mechanisms associated to the evolution of brain networks, we propose a working hypothesis and, provide evidence that a brain dynamical network may evolve based on the maximization of the information flow capacity it can internally handle at each step of its evolution process, i.e. a new inter-neuron connection is established if it leads to a subsequent increase of the information flow capacity of the new brain circuitry.  Our hypothesis is based on the internal neural network dynamics and on a plausible model for brain structure \textit{per se} without the need to resort to probabilistic models based on the external influence in the brain and its response.

We have been able to show that our evolved BDNs present similar synchronization and information flow capacity behaviors with those found for the simulated dynamical networks for the brain structure of the \textit{C.elegans} and humans. Moreover, we show that BDNs evolved with coupling strengths that maximize the information flow capacity are the ones with the smallest spectral graph distance from the BDNs of the \textit{C.elegans} and humans, and that, during the growing process, their information flow capacity increase is related to moderately low amounts of global neural synchronization. This work provides ample evidence that brain networks may grow by maximizing the capacity of information flow they can internally handle, driven by global no Hebbian-like evolution processes, according to which the addition of interconnections between clusters during the evolution process leads to a decrease in the global synchronization level of the BDN. This behavior is accompanied by a similar no Hebbian-like learning process for neurons in some of the clusters and by Hebbian-like processes for neurons in the remaining clusters, leading to an increase in the synchronization level between these neurons during brain network evolution. Effectively, the no Hebbian-like mechanism is \textit{akin} to the unlearning anti-Hebbian mechanism of Crick and Mitchison \cite{Cricketal_1983} that refers to the elimination of unnecessary connections to prevent overload and to render the network more efficient. In our evolution we do not delete links. However, both mechanisms lead to a decrease of synchronization and to more efficient networks, being in our case the evolved networks able to maximize their information flow capacity. Global synchronization takes into consideration the synchronous behavior of all neurons in the BDN whereas local synchronization of neurons in a cluster of the BDNs. Finally, our work shows that optimizing information flow capacity leads to evolved networks that are heterogeneous, in accordance with the line of research in Ref. \cite{Yamaguti2014}, where the authors report on a mathematical model for the evolution of heterogeneous modules in the brain based on the maximization of bidirectional information flow transmission.

\section*{Results}\label{section_results}

In this work we are interested in understanding the relation between synchronization and information flow capacity in BDNs constructed by connecting electrically and chemically, Hindmarsh-Rose (HR) neurons (see Materials and Methods, Subsection Hindmarsh-Rose Neural Model for Brain Dynamics). Their topologies are given by the \textit{C.elegans} and human brain networks, and those that are the result of an evolutionary process that maximizes $I_c$ based on interconnected small-world communities. Our intention is not to strictly model the \textit{C.elegans} and human brain neural dynamics but merely to use the topology of their connectomes to study the corresponding dynamic brain networks and to compare between the results obtained from these cases. In the following, global synchronization measurement (denoted by $\rho$) quantifies the synchronous behavior of all neurons of the network whereas local synchronization (denoted by $\rho_{c_i}$) quantifies the amount of synchronization between ensembles of neurons forming the $i$-th cluster within the BDN. Both are quantified by the order parameter defined in Materials and Methods, Subsection Synchronization Measures in BDNs.

\subsection*{\textit{C.elegans} BDN}\label{subsection_c.elegans_BDN}

\begin{figure}[!ht]
\centering{
\includegraphics[scale=0.05]{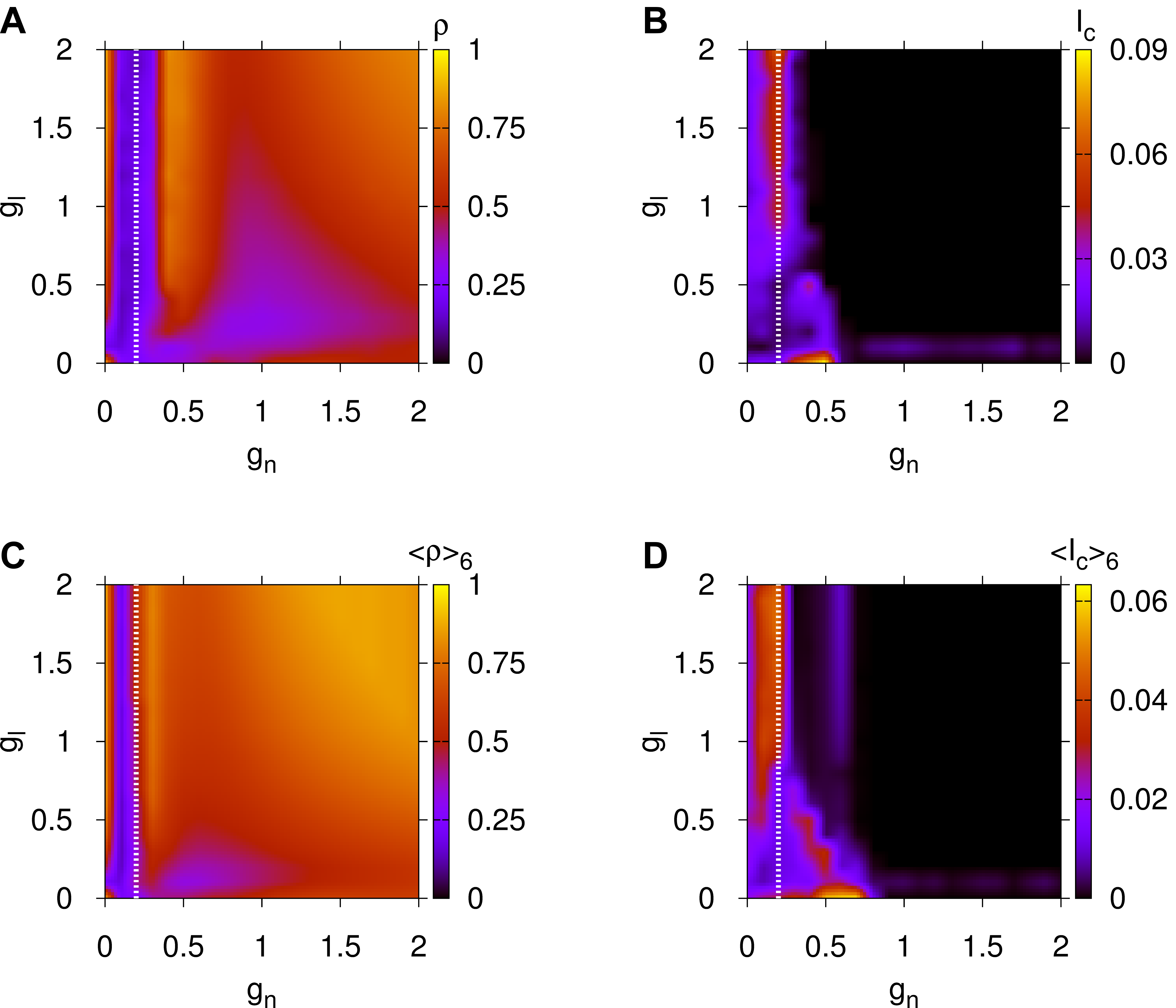}
}
\caption{\textbf{Results for the global synchronization and information flow capacity properties for the \textit{C.elegans} and averaged human BDNs.} Parameter space for the global synchronization $\rho$ in panel (A) and for the upper bound for MIR, $I_c$, in panel (B) for the \textit{C.elegans} BDN. Panels (C) and (D) are similar but for the averaged global synchronization $\langle\rho\rangle_6$ and averaged upper bound for MIR, $\langle I_c\rangle_6$ for the six human BDNs. Here, $g_n$ is the chemical and $g_l$ the electrical coupling of Eqs. \eqref{HR_model_Nneurons}. The regions between the left vertical axes and white dotted lines are replotted in Fig. \ref{fig:zoomin_results_c.elegans_averaged_human_brain} in a finer resolution.}\label{fig:main_results_c.elegans_humans}
\end{figure}

\begin{figure}[!ht]
\centering{
\includegraphics[scale=0.05]{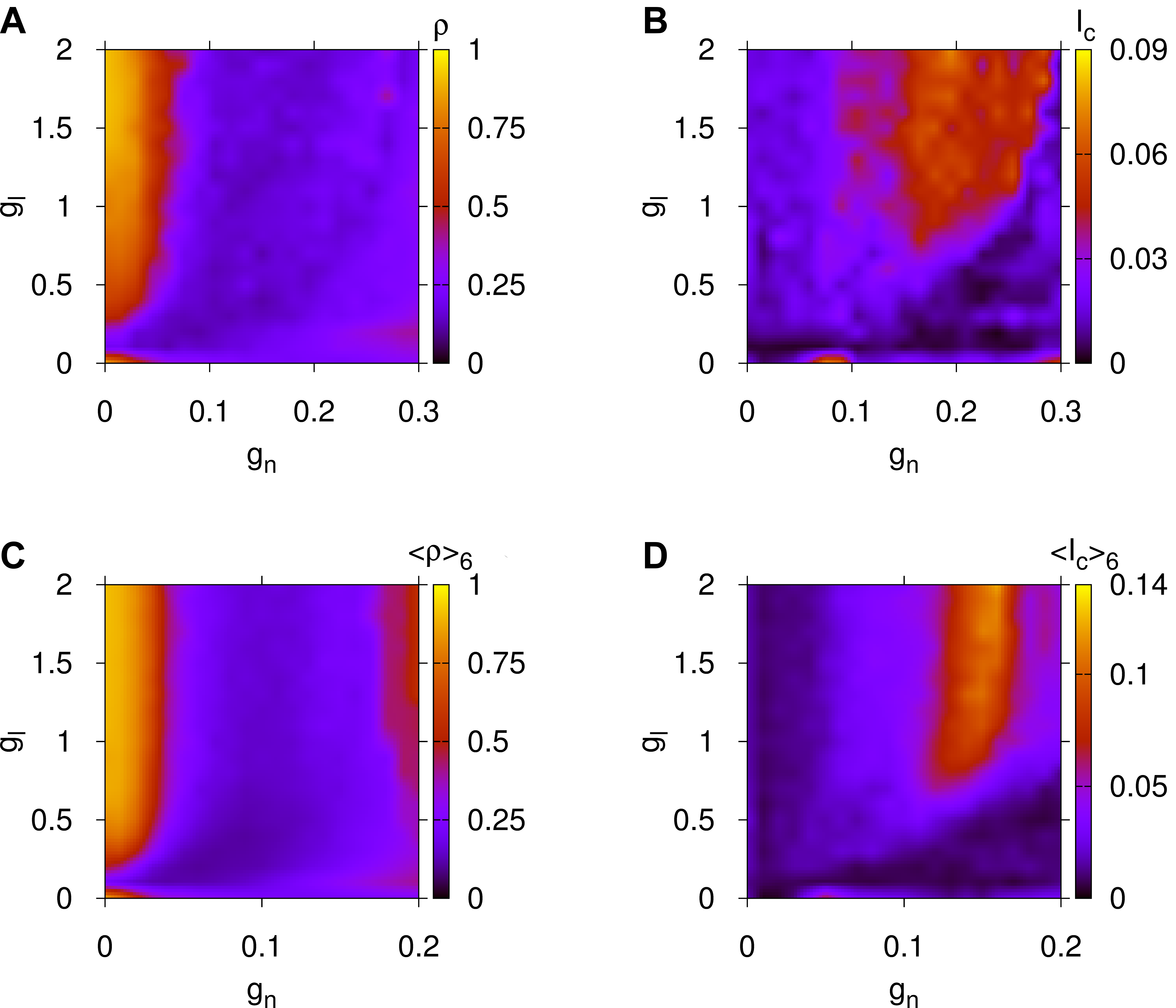}
}
\caption{\textbf{Magnification for the global synchronization and information flow capacity properties of Fig. \ref{fig:main_results_c.elegans_humans}.} Panel (A): Parameter space for $\rho$ and panel (B) for $I_c$ for the \textit{C.elegans} BDN. Panel (C): Similarly for $\langle\rho\rangle_6$ and panel (D) for the averaged upper bound for MIR, $\langle I_c\rangle_6$, of the six human BDNs. Here, $g_n$ is the chemical and $g_l$ the electrical coupling of Eqs. \eqref{HR_model_Nneurons}.}\label{fig:zoomin_results_c.elegans_averaged_human_brain}
\end{figure}

We present in panels (A), (B) of Fig. \ref{fig:main_results_c.elegans_humans} the global synchronization measure $\rho$ and upper bound of information flow $I_c$ for the BDN of the \textit{C.elegans} in the parameter space of chemical coupling $g_n$ in $[0,2]$ and electrical coupling $g_l$ in $[0,2]$. We refer the reader to Materials and Methods, Subsection \textit{C.elegans} Data\ref{subsubsection_c.elegans_data}. A direct comparison between the two panels reveals a number of conclusions for the different parameter space regions. At first, for relatively high chemical and electrical couplings almost full global synchronization can be achieved (yellow and red regions in Fig. \ref{fig:main_results_c.elegans_humans}(A)). For the same region, panel (B) shows an almost absence of capability of information transmission as the upper bound for MIR, $I_c\approx0$ (dark blue region). High levels of global synchronization accompanied by low values of $I_c$ indicate that not only neural activities are similar but also they have very low entropy since both Lyapunov exponents $\lambda_1$, $\lambda_2$ are practically zero leading to their difference $I_c\approx0$ (see Materials and Methods, Subsection Upper Bound for MIR). Secondly, for chemical couplings smaller than 0.3 (i.e. $g_n\in[0,0.3]$) and electrical couplings $g_l\in[0,2]$, we observe a multitude of different functional behaviors: There are regions of high synchronization (red region in panel (A)) and low $I_c$ (blue region in panel (B)) and others with exactly the opposite behavior (i.e. low global synchronization accompanied by high or information flow capacity). These different functional behaviors will become even more evident in Fig. \ref{fig:zoomin_results_c.elegans_averaged_human_brain}, where we plot the regions between the left vertical axes and the white dotted lines of Fig. \ref{fig:main_results_c.elegans_humans} in a finer resolution.

Our findings suggest that for the \textit{C.elegans} BDN, for most of the chemical and electrical couplings, global brain synchronization is roughly speaking inversely related to the information flow capacity, $I_c$. As we shall see in Materials and Methods, Section Similarities between \textit{C.elegans} and human BDNs for a more detailed analysis of the \textit{C.elegans} and human BDNs, high synchronization as depicted by $\rho$ implies small information flow capacity. The relation between $\rho$ and $I_c$ can be better understood in the basis that typically $\rho\propto 1\lambda_2$, where $\lambda_2$ is the second largest Lyapunov exponent of the BDN, for the so-called non-excitatory networks \cite{Baptistaetal2008A,baptista2011complex}. The non-excitatory character is expected to be prominent when the electrical coupling has a dominant contribution to the behavior of the network with respect to the chemical, a situation that promotes global neural synchronization.

\subsection*{Human BDNs}\label{subsection_human_BDN}

We have performed a similar study for the global synchronization $\rho$ and upper bound of information flow $I_c$ for the human brain networks in \textbf{Materials and Methods}, Subsection Human Subjects Data, shown in Fig. \ref{fig:main_results_c.elegans_humans}(C) and (D). Particularly, we first prepared parameter spaces for all six human subject BDNs with the same coupling ranges with those of the first two panels of the \textit{C.elegans} and then computed their average, presented in Fig. \ref{fig:main_results_c.elegans_humans}(C) and (D). The averaged $\rho$ and $I_c$ quantities from the six subjects are indicated by $\langle\rho\rangle_6$ and $\langle I_c\rangle_6$, respectively.

A direct comparison between panels (C) and (D) of Fig. \ref{fig:main_results_c.elegans_humans} for the humans reveals a number of parameter space regions associated to different functional behaviors, similar to those observed for the \textit{C.elegans}. For relatively high chemical and electrical couplings almost full global synchronization is achieved (yellow and red regions in Fig. \ref{fig:main_results_c.elegans_humans}(C)) since $\langle\rho\rangle_6\approx1$. For the same region, Fig. \ref{fig:main_results_c.elegans_humans}(D) shows an almost absence of information flow capacity as $\langle I_c\rangle_6\approx0$, the reason being the same as for the \textit{C.elegans} BDN. For chemical couplings smaller than 0.2 (i.e. $g_n\in[0,0.2]$) and electrical couplings $g_l\in[0,2]$ we observe regions of high synchronization (red regions in Fig. \ref{fig:main_results_c.elegans_humans}(C)) and low $\langle I_c\rangle_6$ (blue region in Fig. \ref{fig:main_results_c.elegans_humans}(D)), as well as others with exactly the opposite property, having low synchronization and large $\langle I_c\rangle_6$.

In order to look deeper into the details of the functional behaviors of these BDNs and, to understand the structural and functional similarities between them, we study in Materials and Methods, Section Similarities between \textit{C.elegans} and human BDNs, zoom-in plots of the previous parameter spaces.

\subsection*{Similarities between \textit{C.elegans} and human BDNs}\label{subsection_similarities_between_c.elegans_and_human_brains}

There has been enormous research devoted on the \textit{C.elegans} worm which has revealed its ability to learn about mechano, chemo and thermosensory inputs and stimuli \cite{Bargmann2006,Kapla1993}. It was also shown that its neural system has the ability to distinguish between tastes, odours or any indication related to the presence or absence of food. It also shows different kinds of learning behavior, such as associative (classical conditioning and differential classical conditioning), and non-associative forms of learning, such as habituation and dishabituation \cite{Gallyetal2003}. These properties are reminiscent of the human brain ability to adapt to different stimuli and environments.

In Fig. \ref{fig:zoomin_results_c.elegans_averaged_human_brain}, we present a finer resolution version of the parameter spaces of Fig. \ref{fig:main_results_c.elegans_humans} for the \textit{C.elegans} and humans. They allow us to reveal the extraordinary similarity on the functional level, the global synchronization and $I_c$ patterns between the \textit{C.elegans} and human BDNs. In these plots, wherever we observe high synchronization as evidenced by orange and red regions in panel (A) for the \textit{C.elegans} and (C) for the human brain network, the upper bound for MIR, $I_c$, that stands for the internal information flow capacity of the BDN, is small (blue region) and vice-versa.

Since the \textit{C.elegans} brain connectivity network is about four times smaller in size than the human BDNs \cite{Hagmannetal2008} studied here, we should expect that the global synchronization and upper bound for MIR patterns could occur for different ranges of chemical and electrical couplings as compared to those of Fig. \ref{fig:main_results_c.elegans_humans} (see for example Ref. \cite{Baptistaetal2010}). Therefore, a rescaling of the coupling strengths was employed to allow for both BDNs to have the possibility to produce equivalent dynamical behaviors. This rescaling is described in Materials and Methods, Subsection Rescaling of Chemical and Electrical Couplings for Parameter Spaces of Networks with Different Eigenvalue Spectra. It is worth noting that there is an optimal coupling range for both BDNs that allows for large information flow capacity in the brain networks (orange and yellow regions in panels (B), (D) of Fig. \ref{fig:zoomin_results_c.elegans_averaged_human_brain}), a coupling range that promotes moderately low global synchronization!

\subsection*{Functional Properties of the Model for Brain Network Evolution}\label{subsection_functional_properties_of_the_model_for_Brain_evolution}

In Materials and Methods, Subsection A Model for Brain Network Evolution Based on the Maximization of Information Flow Capacity, we propose an artificial brain network evolution model that presents important structural and functional properties of the BDNs of the \textit{C.elegans} and humans. It is based on the combined effect of chemical and electrical synapses and, on a topology reminiscent of interconnected brain communities found in these BDNs. We use chemical synapses for the communication of neurons of different clusters (inter-cluster connections) and electrical for the communication of neurons within each cluster (intra-cluster connections).

Here, we compare the functional properties of this brain network evolution model with those of the \textit{C.elegans} and humans. By functional we mean the properties of the dynamics of the BDN such as local, global synchronization and information flow capacity. Particularly, we report on the similarities we found for the functional properties of the BDNs of the \textit{C.elegans} and humans, and for those of the proposed model for brain network evolution.

\begin{figure}[!ht]
\centering{
\includegraphics[scale=0.033]{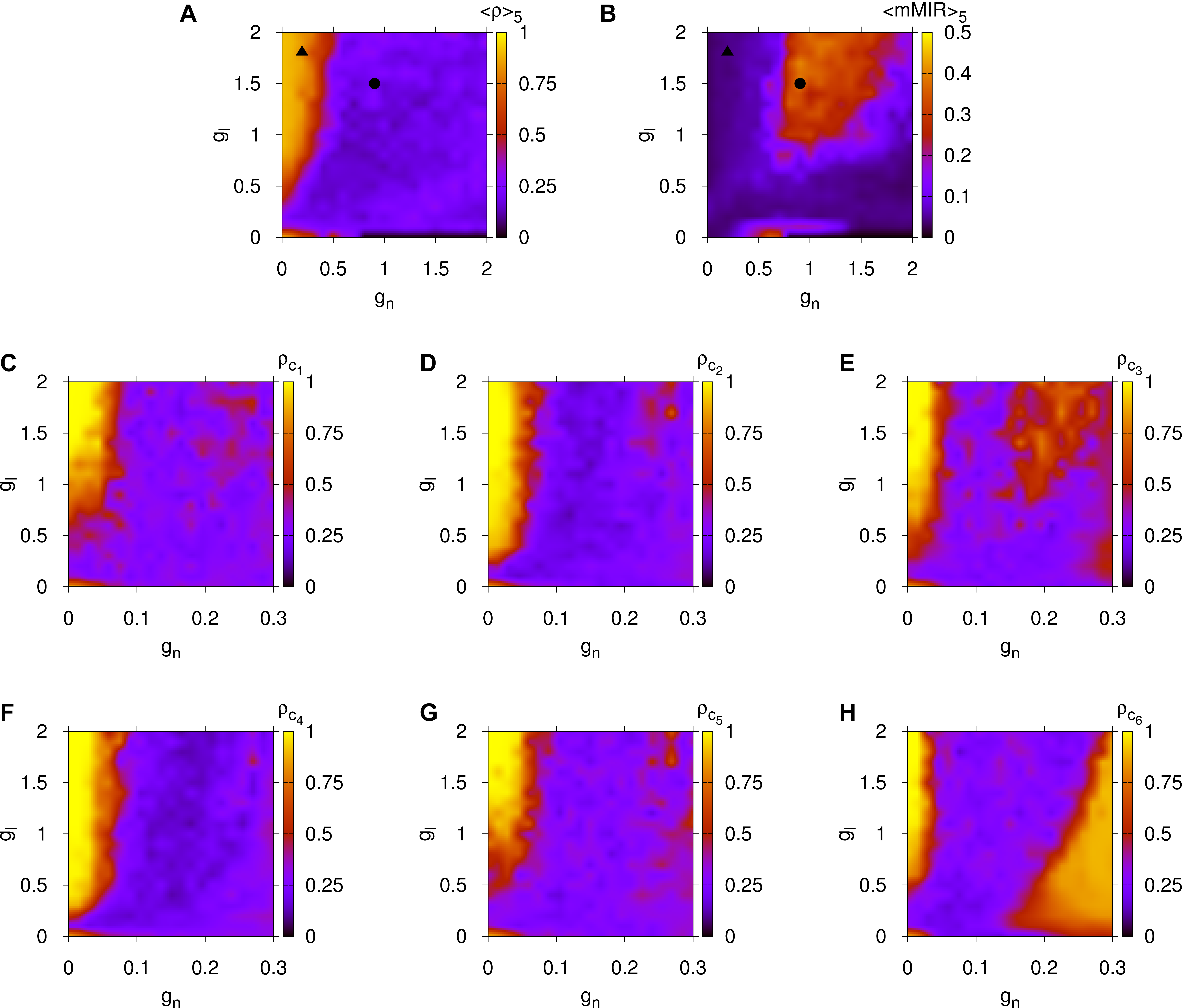}
}
\caption{\textbf{Results for the global and local synchronization, and information flow capacity properties for the evolved networks of Materials and Methods, Subsection A Model for Brain Network Evolution Based on the Maximization of Information Flow Capacity.} Panel (A): Parameter space for the synchronization $\langle\rho\rangle_5$. Panel (B): Parameter space for the averaged upper bound for MIR, $\langle\mbox{mMIR}\rangle_5$, from the five realizations of a network of 60 neurons with six, equally sized, small-world clusters. Case $\mathcal{A}$ of high synchronization and low information flow capacity is denoted by $\blacktriangle$ and case $\mathcal{B}$ of low synchronization and high information flow capacity by $\bullet$. Panels (C) to (H) are plots for the local synchronization $\rho_{c_i}$ of the six communities of the \textit{C.elegans} brain network. To be compared with panel (A). Here, $g_n$ is the chemical and $g_l$ the electrical coupling of Eqs. \eqref{HR_model_Nneurons}.}\label{fig:main_results_n60c6}
\end{figure}

To support further the validity of the presented results for the model for brain network evolution and to show its independence on the particular initial small-world cluster configuration, we computed its parameter spaces averaging over the functional measurements obtained from five different initial small-world cluster configurations (for a discussion about the creation of small-world networks or clusters see Materials and Methods, Subsection Analysis of Networks and Communities). In all cases, we used the same evolution process as described in Materials and Methods, Subsection A Model for Brain Network Evolution Based on the Maximization of Information Flow Capacity. The results of this study are shown in Fig. \ref{fig:main_results_n60c6}, where $\langle\rho\rangle_5$ stands for the average of the global synchronization of the evolved BDNs for the five realizations and $\langle\mbox{mMIR}\rangle_5$ for the average maximal value of the MIR for the same realizations. We first provide evidence in panel (A) and (C) to (H) that the finally evolved BDN of small-world clusters captures similar local and global synchronization properties to those of the \textit{C.elegans} BDN. The local synchronization $\rho_{c_i}$ of the $i$-th community (Fig. \ref{fig:main_results_n60c6}(C) - (H)) also reproduces similarly the global synchronization patterns of Figs. \ref{fig:zoomin_results_c.elegans_averaged_human_brain}(A) and (C) for the \textit{C.elegans} and human BDNs. Comparing these panels, we conclude that the averaged synchronization measure $\langle\rho\rangle_5$ of the brain network evolution model (Fig. \ref{fig:main_results_n60c6}(A)) attains almost similar values in the same coupling regions to those of the \textit{C.elegans} in Fig. \ref{fig:zoomin_results_c.elegans_averaged_human_brain}(A). The different ranges on the horizontal axes of the chemical coupling strength $g_n$ can be explained by the fact that the finally evolved network consists of 60 neurons whereas the \textit{C.elegans} of 277 neurons (for more details see Materials and Methods, Subsection Rescaling of Chemical and Electrical Couplings for Parameter Spaces of Networks with Different Eigenvalue Spectra).

It is remarkable that the finally evolved BDN exhibits almost identical synchronization and information flow capacity features as the BDN of the \textit{C.elegans} and humans for almost all coupling ranges considered. Particularly, focusing on panels (A), (B) (for the \textit{C.elegans}) and (C), (D) (for the humans) of Fig. \ref{fig:zoomin_results_c.elegans_averaged_human_brain} and, on (A), (B) of Fig. \ref{fig:main_results_n60c6} (for the brain network evolution model), we observe an almost identical pattern of global synchronization and information flow capacity as depicted by $I_c$ and its averages. Again, here we have used the rescaling in Materials and Methods, Subsection Rescaling of Chemical and Electrical Couplings for Parameter Spaces of Networks with Different Eigenvalue Spectra, to create networks that can potentially reproduce similar dynamical behaviors for the finally evolved BDNs, in agreement with those identified for the \textit{C.elegans} and human BDNs earlier.

The upper bound for MIR, $I_c$, depends on various factors, such as network topology and connectivity patterns, coupling strengths and types (chemical, electrical), synchronicity, etc. The values of $I_c$, the information flow capacity, can be comparable (or different) for networks with different topologies. This is due to the fact that $I_c$ is a function of the chemical and electrical coupling strengths. The surprising fact is that as we evolve a network with an initial small-world network configuration, by maximizing information flow capacity, the final network exhibits not only similar topological (structural network characteristics) but also similar functional or behavioral (synchronization and upper bound for MIR) properties as those found for the brain dynamical networks of the \textit{C.elegans} and human subjects (see Materials and Methods, Subsection Spectral Similarity of \textit{C.elegans} and Human Brain Networks with those of the Model for Brain Network Evolution).

For completeness, in Figure S2, we present a similar analysis to the one of Fig. \ref{fig:main_results_n60c6} based on Erd\H{o}s-R\'enyi random networks (panels (A), (B)), scale-free (Barab\'asi-Albert) (panels (C), (D)) and star topologies perturbed by $20\%$ for the clusters of the model for brain evolution of 60 neurons and 6 small-world clusters (panels (E), (F)) and found out that the evolution model fails to capture similar functional (local and global synchronization and, information flow capacity patterns in the parameter spaces) as the same model equipped with small-world topologies for its clusters (see Fig. \ref{fig:main_results_n60c6}). Here, we are interested in studying and proposing a model for brain network evolution that is able to reproduce not only similar functional properties such as information flow capacity and, local and global synchronization properties, but also importantly to reproduce similar structural properties for the finally evolved full brain network and of its clusters. Based on our results so far, we show next that the initial cluster configuration able to fulfil both requirements is the small-world cluster topology.

\subsection*{Spectral Similarity of \textit{C.elegans} and Human Brain Networks with those of the Model for Brain Network Evolution}\label{subsubsection_comparison_between_spectra_of_realistic_brain_networks_and_proposed_evolution_model}

\begin{figure}[!h]
\centering{
\includegraphics[scale=0.038]{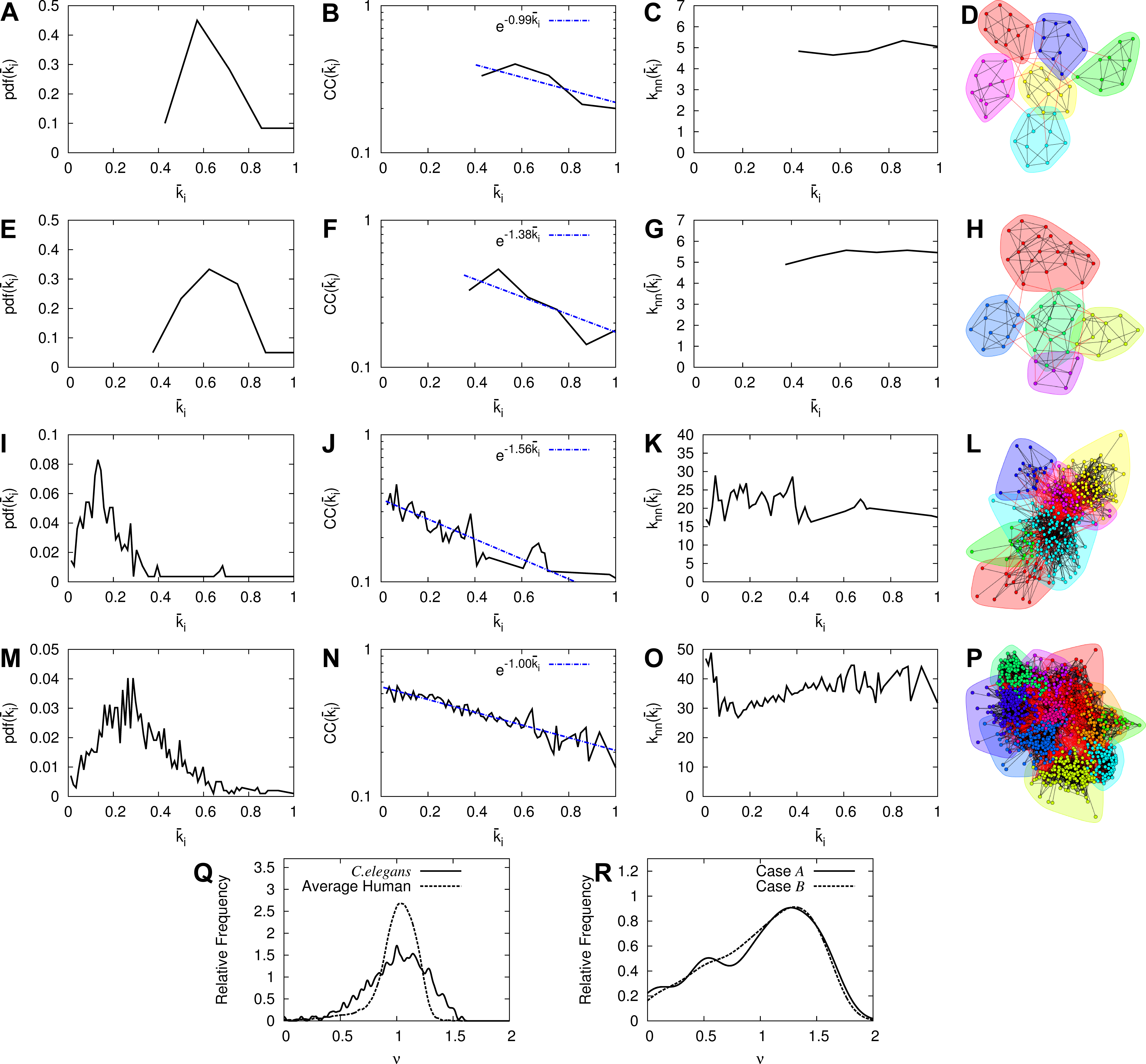}
}
\caption{\textbf{Structural properties and normalized Laplacian spectra of the brain networks considered in this study.} Panels (A) to (D): Plot of the pdf of the normalized degrees $\bar{k}_i$ (panel (A)), plot of the clustering coefficient $CC(\bar{k}_i)$ (panel (B)), plot of the average normalized degree $k_{nn}(\bar{k}_i)$ of the neighbors of nodes with normalized degree $\bar{k}_i$ (panel (C)) and the network with its distinct clusters and communities given by different colors for the case $\mathcal{A}$ of high synchronization and low mMIR of the model for brain network evolution of 60 neurons and 6 clusters. Panels (E) to (H): Same as in panels (A) to (D) but for the case $\mathcal{B}$ of low synchronization and high mMIR of the same model. Panels (I) to (L): Same as in panels (A) to (D) but for the \textit{C.elegans} brain network. Panels (M) to (P): Same as in panels (A) to (D) but for the human subject A$_1$ brain network. In the plots of the second column, we show with blue dashed lines the exponential dependence of CC($\bar{k}_i$) to $\bar{k}_i$ to guide the eye, where $\bar{k}_i$ is the normalized degree. It is defined as $\bar{k}_i=k_i/k_{max}$, where $k_i$ is the node degree and $k_{max}$ is the largest node degree in the network. For the first row, $k_{max}=7$, for the second $k_{max}=8$, for the third $k_{max}=76$ and for the fourth, $k_{max}=87$. Panel (Q): Normalized Laplacian spectra of the brain network of \textit{C.elegans} (solid curve) and of the averaged over the six human subjects (dashed curve). Panel (R): Similarly for the brain network of case $\mathcal{A}$ of high synchronization and low information flow capacity (solid curve) and, for case $\mathcal{B}$ of low synchronization and high information flow capacity (dashed curve) of the model for brain network evolution.}
\label{fig:statistical_analysis_brain_networks}
\end{figure}

The discussion about the results of Fig. \ref{fig:statistical_analysis_brain_networks} in Materials and Methods suggest that the evolution process of a basic small-world clustered network is capable of generating an evolved one with similar structural properties with those for the \textit{C.elegans} and human BDNs (see Materials and Methods, Subsection Structural Properties of the Model for Brain Network Evolution). The structural similarity to the human brain network is even more remarkable if the couplings of the network to be evolved are within the range that promotes high levels of information flow capacity and low neural synchronization, a prominent example of which is case $\mathcal{B}$ (for a definition and discussion about cases $\mathcal{A}$, $\mathcal{B}$, see Materials and Methods, Subsection Brain Network Evolution Promotes Global no Hebbian-like and, Local Hebbian-like and no Hebbian-like Evolution Learning).

\begin{figure}[!ht]
\centering{
\includegraphics[scale=0.04]{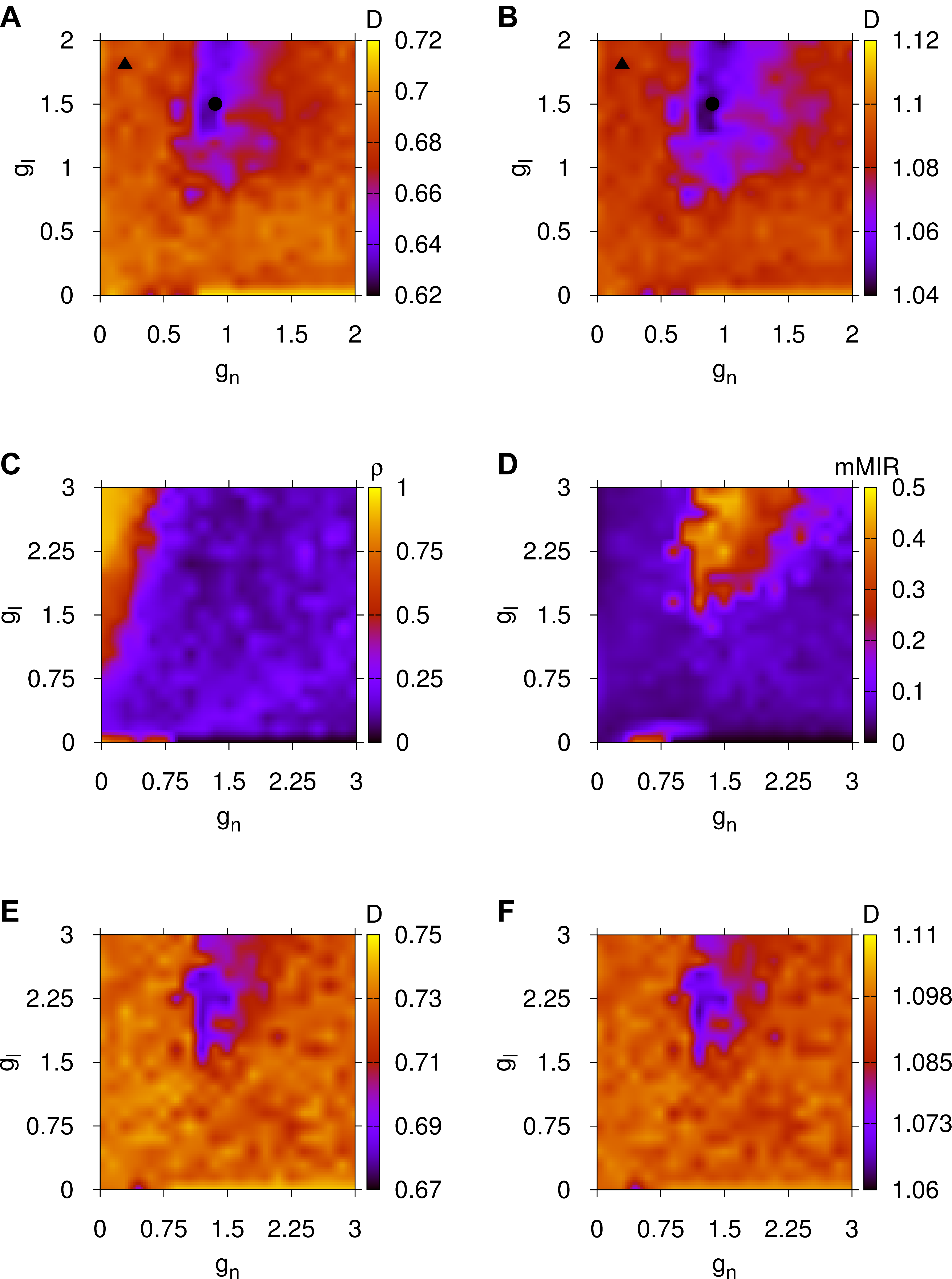}
}
\caption{\textbf{Spectral distances between the network topology of the \textit{C.elegans}, humans and evolved networks and, global synchronization and information flow capacity properties for an evolved BDN of 120 neurons.} Panel (A): Parameter space for the spectral distance $D$ between the \textit{C.elegans} brain network and the averaged model for brain network evolution of 60 neurons and six small-world clusters and, panel (B) similarly for the spectral distance between the averaged brain network of the six human subjects and the same averaged network created by our brain network evolution. $\blacktriangle$ denotes case $\mathcal{A}$ and $\bullet$ case $\mathcal{B}$, both explained in Materials and Methods, Subsection Brain Network Evolution Promotes Global no Hebbian-like and, Local Hebbian-like and no Hebbian-like Evolution Learning Processes. Both panels to be compared with Fig. \ref{fig:main_results_n60c6}(B). Panel (C): Parameter space for the synchronization $\rho$ and panel (D) for the mMIR of evolved networks using our brain network evolution process with 120 neurons. Panels (E), (F) are similar to (A), (B) for the spectral distance between the \textit{C.elegans}, averaged brain network of the six human subjects and the model for brain network evolution of 120 neurons. Panels (E), (F) to be compared with panel (D). Here, $g_n$ is the chemical and $g_l$ the electrical coupling of Eqs. \eqref{HR_model_Nneurons}.}\label{fig:spectral_distances}
\end{figure}

Here we study how close the normalized Laplacian spectral plots of the networks considered in Fig. \ref{fig:statistical_analysis_brain_networks}(Q) are from those of evolved BDNs. Particularly, we examined the spectral similarity by comparing spectral plots and computing their average Euclidean distance following Ref. \cite{Lange_2014}. We refer the reader to Materials and Methods, Subsection Normalized Laplacian Spectra, for the details of our computations. We compared the normalized Laplacian spectral plot of the \textit{C.elegans} and of the six human brain networks with the spectral plots of all finally evolved networks of the five realizations used to compute the averaged parameter spaces of the model for brain network evolution of Fig. \ref{fig:main_results_n60c6}(A), (B). We provide additional support by demonstrating in the last four panels of Fig. \ref{fig:spectral_distances} results from a similar comparison between the \textit{C.elegans} and humans with a double-sized version of 120 neurons of the model for brain network evolution of the Materials and Methods, Subsection A Model for Brain Network Evolution Based on the Maximization of Information Flow Capacity. The normalized Laplacian spectra for the \textit{C.elegans} and for the averaged over the six human BDNs are shown in Fig. \ref{fig:statistical_analysis_brain_networks}(Q). As it can be seen, both spectra share some common features: Both show a left-skewed distribution in which the largest eigenvalue is closer to one in agreement with results in Ref. \cite{Lange_2014}. Also, the distributions show peaks around one and the eigenvalues are scattered at the beginning of the spectra, suggesting similarities in their community structure, reminiscent of their small-worldness.

Although the spectral plots of both cases of the model for brain network evolution shown in Fig. \ref{fig:statistical_analysis_brain_networks}(R) do not exhibit all properties of the spectral plots of the \textit{C.elegans} and humans of Fig. \ref{fig:statistical_analysis_brain_networks}(Q), maybe due to the considerably smaller size of the former networks, they do exhibit interesting similarities: They are both left-skewed distributions with a peak around 1.3, being closer to 1 than to 2. We also observe low relative frequency eigenvalues at the beginning of both spectra. Both spectral properties suggest similarities in their community structure \cite{Lange_2014} as well (i.e. their small-worldness). This is in accordance with the spectral plots of the averaged human and \textit{C.elegans} brain networks of Fig. \ref{fig:statistical_analysis_brain_networks}(Q). We argue that this similarity comes from the small-worldness of the communities. The close relation between the normalized Laplacian spectra of Fig. \ref{fig:statistical_analysis_brain_networks} suggests the existence of common underlying structural properties of the neural networks of the \textit{C.elegans}, the humans and the model for brain network evolution.

We measured the similarity between the spectral plots of the \textit{C.elegans} and the averaged human brain network with the averaged model for brain network evolution and, plot in panels (A), (B) of Fig. \ref{fig:spectral_distances} their spectral distance $D$ for different chemical and electrical coupling ranges. In this framework, the closer $D$ is to zero, the closer structurally the compared networks are. In particular, panel (A) is the parameter space for the spectral distance between the \textit{C.elegans} and the averaged model for brain network evolution and, panel (B) is a similar plot for the spectral distance between the averaged human brain network and the same model for brain network evolution. We pinpoint case $\mathcal{A}$ by a $\blacktriangle$ and case $\mathcal{B}$ by $\bullet$.

Such results allow one to draw interesting relations between structure and function of the proposed model for brain network evolution with the structural properties of the brain networks of \textit{C.elegans} and humans. Panels (A) and (B) of Fig. \ref{fig:spectral_distances} already reveals that the smallest mean spectral distance happens for the pair of chemical and electrical couplings that gives rise to BDNs that present small amount of synchronization and high information flow capacity, in other words to cases such as $\mathcal{B}$. In contrast, one of the largest mean spectral distances was found for case $\mathcal{A}$ that promotes high amount of neural synchronization and small information flow capacity in the brain network!

\section*{Materials and Methods}\label{online_methods}

\subsection*{Preparation of Data for the Study of BDNs}\label{subsection_preparation_of_data_for_the_study_of_brain_network}

\paragraph*{\textit{C.elegans} Data}\label{subsubsection_c.elegans_data}

\textit{C.elegans} is the most commonly used model organism for neural network studies. It is a 1mm long soil worm with a simple nervous system that can be represented at first by 302 neurons and about 7000 synapses \cite{Gallyetal2003}. Its nervous system is divided into two distinct and independent nervous systems: A large somatic nervous system with 282 neurons and a small pharyngeal nervous system with only 20 neurons. Since neurons CANL/R and VC06 do not make synapses with other neurons (i.e. they are isolated), they can be removed and a reduced connectivity brain matrix of 279 somatic neurons can be produced. We use in our study the connectome of the large somatic nervous system found in Ref. \cite{cmtkdataset} that consists of 277 neurons. We decided to use the undirected version of this adjacency matrix as we are not concerned with the directionality of the information flow. We simulate the dynamics of each ``neuron'' by a single HR neuron system given in Eq. \eqref{HR_model_1neuron} and couple them by the corresponding adjacency matrix obtained from the brain connectivity of the \textit{C.elegans} using Eqs. \eqref{HR_model_Nneurons}.

We study the \textit{C.elegans} nervous system in order to understand the human nervous system. The reason is that both human and \textit{C.elegans} nervous systems consist of neurons and the communication or flow of information is passing through synapses that use neurotransmitters to perform brain activity. Many of these neurotransmitters are common in humans and \textit{C.elegans} such as Glutamate, GABA, Acetylcholine and Dopamine. The genome of the \textit{C.elegans} is almost 30 times smaller than that of humans but still, encodes almost 22000 proteins. Moreover, it is almost 35\% similar to that of humans \cite{Blumenthal1996,mcdgicelegans}.

\paragraph*{Human Subjects Data}\label{subsubsection_human_subjects_data}
 
The data for the analysis of the human connectome were based on Refs. \cite{Hagmannetal2008,cmtkdataset}. The authors report on results based on the study of five different subjects (right handed males aged between 24 and 32) coded as A, B, C, D and E, where the first one was examined twice giving connectomes A$_1$, A$_2$ with the second examination performed several days after the first one which yielded a highly consistent regional adjacency matrix A$_2$ to A$_1$. In our study, we make use of all six adjacency matrices (referring thereafter to six human subject BDNs) to prepare averaged quantities for global synchronization and upper bound for MIR. The diffusion spectrum imaging technique was then employed to retrieve high-resolution connection matrices for all subjects. The cortical regions of their brains were then further divided into 66 clear anatomical regions and these were individually subdivided into smaller regions of interest (of size 1.5cm$^2$) which finally resulted in 998 parts. These parts cover the entire cortices of both hemispheres but do not include subcortical nodes and connections \cite{Hagmannetal2008}. Thus, the original adjacency matrices obtained for brain circuitry had 998 neural ensembles for all subjects. However, during the analysis of the adjacency matrices of these subjects, we found out that all these brain networks were disconnected and different numbers of isolated neural ensembles (nodes) were identified for each connectome. We decided to remove the isolated nodes as they were not connected to other neural ensembles of the connectome, and ended up with 994 for subject A$_1$, 987 for A$_2$, 980 for B, 996 for C and D, and 992 for E. We also decided to use the undirected versions of the adjacency matrices for the same reason as for the \textit{C.elegans}. We simulated each neural ensemble by a single HR neuron given in Eq. \eqref{HR_model_1neuron} and coupled them by the corresponding adjacency matrix obtained from the brain connectivity using Eqs. \eqref{HR_model_Nneurons}.

\subsection*{Hindmarsh-Rose Neural Model for Brain Dynamics}\label{subsection_Hindmarsh-Rose_Model_for_Brain_Dynamics}

The complexity of the circuitry of the nervous system of the human brain is still a big challenge to be resolved as it contains about 86 billion neurons and thousands times more synapses \cite{Azevedo2009}. A synapse is a junction between two neurons and it is a mean through which neurons communicate with each other. There are electrical and chemical synapses: An electrical synapse is a physical connection between two neurons which allows electrons to pass through neurons by a very small gap between nerve cells. Electrical synapses are bidirectional and of a local character, happening between neurons whose cells are close. They are believed to contribute to the regulation of synchronization in the brain network. In contrast, chemical synapses are special junctions through which the axon of the pre-synaptic neuron comes close to the post-synaptic cell membrane of another neuron or non-neural cell. In Ref. \cite{Varshneyetal2011}, the authors report on the self-consistent gap junctions and chemical synapses in the connectome of the \textit{C.elegans}. In our work, we use both kinds of synapses.

Following Ref. \cite{Baptistaetal2010}, we endow the nodes (i.e. neurons for the \textit{c.elegans} and neural ensembles for the humans) of the networks with Hindmarsh-Rose brain dynamics \cite{Hindmarshetal1984}:
\begin{eqnarray}\label{HR_model_1neuron}
 \dot{p}=q-ap^3+bp^2-n+I_{\mbox{ext}}\nonumber,\\
 \dot{q}=c-dp^2-q\nonumber,\\
 \dot{n}=r[s(p-p_0)-n],
 \end{eqnarray}
where $p$ is the membrane potential, $q$ is associated with the fast current, $Na^{+}$ or $K^{+}$, and $n$ with the slow current, for example $Ca^{2+}$. The rest of the parameters are defined as $a=1$, $b=3$, $c=1$, $d=5$, $s=4$, $p_0=-1.6$ and $I_{\mbox{ext}}=3.25$ for which the system exhibits a multi-scale chaotic behavior characterized as spike bursting. $r$ modulates the slow dynamics of the system and was set to 0.005 so that each neuron or neural ensemble be chaotic. For these parameters, the HR model enables the spiking-bursting behavior of the membrane potential observed in experiments made with a single neuron \textit{in vitro}. It is also a relatively simple model that provides a good qualitative description of the many different patterns empirically observed in neural activity.

We couple the HR system to create an undirected BDN of $N_n$ neurons connected simultaneously by electrical (linear diffusive coupling) and chemical (nonlinear coupling) synapses:
\begin{eqnarray}\label{HR_model_Nneurons}
 \dot{p}_i=q_i-a p_i^3+bp_i^2-n_i+I_{\mbox{ext}}-g_n(p_i-V_{\mbox{syn}})\sum_{j=1}^{N_n}\boldsymbol{B}_{ij}S(p_j)-g_l\sum_{j=1}^{N_n}\boldsymbol{G}_{ij}H(p_j)\nonumber,\\
 \dot{q}_i=c-dp_i^2-q_i\nonumber,\\
 \dot{n}_i=r[s(p_i-p_0)-n_i],\nonumber\\
 \dot{\phi}_i=\frac{\dot{q}_i p_i-\dot{p}_i q_i}{p_i^2+q_i^2},\;i=1,\ldots,N_n.
\end{eqnarray}
In our study, $\dot{\phi}_i$ is the instantaneous angular frequency of the $i$-th neuron and $\phi_i$ is the phase defined by the fast variables $(p_i,q_i)$ of the $i$-th neuron. We consider $H(p_i)=p_i$ and:
\begin{equation}\label{S_p_j}
 S(p_j)=\frac{1}{1+e^{-\lambda(p_j-\theta_{\mbox{syn}})}},
\end{equation}
with $\theta_{\mbox{syn}}=-0.25$, $\lambda=10$, and $V_{\mbox{syn}}=2$ to create excitatory BDNs. Equation \eqref{S_p_j} is a sigmoidal function that acts as a continuous mechanism for the activation and deactivation of the chemical synapses and, also allows for analytical calculations of the synchronous modes and synchronization manifolds of the coupled system of Eqs. \eqref{HR_model_Nneurons} \cite{Baptistaetal2010}. In Eqs. \eqref{HR_model_Nneurons}, $g_n$ is the coupling strength associated to the chemical synapses and $g_l$ to the electrical synapses. For the chosen parameters, we have $|p_i|<2$ and that $(p_i-V_{\mbox{syn}})$ is always negative for excitatory networks. If two neurons are connected under an excitatory synapse then, when the presynaptic neuron spikes, it induces the postsynaptic neuron to spike. We adopt only excitatory chemical synapses in our analysis. We use as initial conditions for each neuron $i$: $p_i=-1.30784489+\eta^r_i$, $q_i=-7.32183132+\eta^r_i$, $n_i=3.35299859+\eta^r_i$ and $\phi_i=0$, where $\eta^r_i$ is a uniformly distributed random number in $[0,0.5]$ for all $i=1,\ldots,N_n$, following Ref. \cite{Baptistaetal2010}.  These initial conditions place the trajectory quickly in the attractor of the dynamics and thus, there is less need to consider longer transients.

$\boldsymbol{G}_{ij}$ accounts for the way neurons are electrically (diffusively) coupled and it is a Laplacian matrix (i.e. $\boldsymbol{G}_{ij}=\boldsymbol{K}_{ij}-\boldsymbol{A}_{ij}$, where $\boldsymbol{A}$ is the binary adjacency matrix of the electrical connections and $\boldsymbol{K}$ is the degree identity matrix based on $\boldsymbol{A}$), and so $\sum_{j=1}^{N_n}\boldsymbol{G}_{ij}=0$. By a binary adjacency matrix, we mean an adjacency matrix with entries either 0 (no connection) or 1 (connection). $\boldsymbol{B}_{ij}$ is a binary adjacency matrix and describes how the neurons are chemically connected and therefore its diagonal elements are equal to 0, giving thus $\sum_{j=1}^{N_n}\boldsymbol{B}_{ij}=k_i$, where $k_i$ is the degree of the $i$-th neuron, i.e. it represents the number of chemical links that neuron $i$ receives from all other $j$ neurons in the network. A positive (i.e. 1) off-diagonal value in both matrices $\boldsymbol{A}, \boldsymbol{B}$ in row $i$ and column $j$ means that neuron $i$ perturbs neuron $j$ with an intensity given by $g_l\boldsymbol{G}_{ij}$ (electrical diffusive coupling) or $g_n\boldsymbol{B}_{ij}$ (chemical excitatory coupling), respectively. Therefore, the binary adjacency matrices $\boldsymbol{C}$ of the BDNs considered in this work are given by:
\begin{equation}\nonumber
\boldsymbol{C}=\boldsymbol{A}+\boldsymbol{B}.
\end{equation}

\subsection*{Numerical Simulations Details}\label{subsection_numerical_simulations_details}

We numerically integrated Eqs. \eqref{HR_model_Nneurons} in Fortran 90 using the Euler integration method (order one) with time step $\delta t=0.01$. We decided to do so to reduce the numerical complexity and CPU time of the required simulations to feasible levels as a preliminary comparison of trajectories computed for the same parameters (i.e. $\delta t$, initial conditions, etc.) with integration methods of order 2, 3 and 4, revealed similar results.

We evolve the dynamics of the brain networks and calculate their two largest Lyapunov exponents $\lambda_1$, $\lambda_2$ for the estimation of the upper bound for MIR, $I_c$. We use the well-known method of Refs. \cite{Benettin1980a,Benettin1980b} to compute the Lyapunov exponents needed for the estimation of the upper bound $I_c$ for MIR (see also Materials and Methods, Subsection Upper Bound for MIR). The numerical integration of the HR system of Eqs. \eqref{HR_model_Nneurons} for the \textit{C.elegans} and human BDNs was performed for the final integration time $t_f=5000$ and the computation of the different quantities such as the order parameter $\rho$ of Materials and Methods, Subsection Synchronization Measures in BDNs and the Lyapunov exponents, starts after the transient time $t_t=300$ to make sure that orbits converged to the attractor of the dynamics. The same parameters for the model of brain network evolution of Materials and Methods, Subsection A Model for Brain Network Evolution Based on the Maximization of Information Flow Capacity were set to $t_f=2500$ and $t_t=300$ to reduce the numerical complexity and CPU time to feasible levels, retaining similar results. We have been careful to check that using exactly the same values as for the \textit{C.elegans} and humans, the conclusions were practically the same.

\subsection*{A Model for Brain Network Evolution Based on the Maximization of Information Flow Capacity}\label{subsection_an_information_based_model_for_brain_evolution}

In this work we propose an artificial evolution model for brain network connectivity that captures important structural and functional properties of the BDNs of the \textit{C.elegans} and humans. Our idea is reminiscent of modular processors that are sufficiently isolated and dynamically differentiated to achieve independent computations, but also globally connected to be integrated in coherent functions \cite{Meunieretal2001}.

The model is based on the consideration of the combined effect of chemical and electrical synapses between neurons based on a topology reminiscent of interconnected brain network communities found in the BDNs of \textit{C.elegans} and humans. In this study, we consider chemical synapses solely for the communication of neurons of different clusters of the network (inter-cluster connections) and electrical synapses for the communication of neurons within each cluster (intra-cluster connections). This idea comes from the biological local and non-local nature of these connections.

Particularly, we consider a starting network topology for brain network evolution, where $N_c$ clusters of electrically coupled neurons are connected in a closed ring as shown in Figure S1 of the Supporting Information. We endow each cluster with a small-world topology \cite{Wattsetal1998} as this is what we found to be more plausible to happen on the brain networks of the \textit{C.elegans} and humans (see Subsection Analysis of Networks and Communities\ref{subsection_analysis_of_communities} in Materials and Methods). We also use in our evolution model, for simplicity but this is not mandatory, clusters of the same number of neurons. We denote the total amount of neurons in the network by $N_n$. Each small-world cluster is connected by an inter-cluster connection with its two nearest neighbour clusters by only chemical excitatory connections. In Figure S1 of the Supporting Information, one can see such an example of a small-world network topology that comprises $N_c=6$ clusters and $N_n=60$ neurons, where the red links denote the chemical inter-cluster connections and black the electrical intra-cluster connections. For this model network, we subsequently compute the two largest Lyapunov exponents $\lambda_1,\lambda_2$ of the BDNs following Refs. \cite{Benettin1980a,Benettin1980b} to estimate the upper bound $I_c=\lambda_1-\lambda_2$ for the MIR of the network, i.e. the maximum amount of information per time unit that can be exchanged between the neurons of the basic (not yet evolved) network, \textit{aka} its information flow capacity (for the details see Subsection Upper Bound for MIR\ref{subsection_Upper_Bound_for_MIR} in Materials and Methods).

We then evolve the starting network, such as the one in Figure S1, by adding new chemical excitatory inter-cluster connections to simulate the creation of new chemical synapses between neurons of different clusters. The electrical connections, topology and the values of the chemical and electrical coupling strengths are not modified during the evolution process. We adopt the following evolutionary rule to imitate brain plasticity \cite{Gerstneretal2008}: If the newly added inter-cluster chemical connection leads to an increase of $I_c$ prior to the addition, the new synapse is retained. If, instead, it is found not to increase $I_c$ then it is deleted from the network and the random search for another one starts, being this procedure iterative. We choose the nodes of the different small-world clusters so that to simulate the addition of new inter-cluster connections in a random fashion (i.e. the candidate links are randomly chosen from a uniform distribution).

The iterative procedure is repeated until the maximum number of possible pairs of neurons from different clusters is exhausted. We denote by mMIR the value of $I_c$ of the finally evolved BDN which is always bigger or equal than the $I_c$ of the starting BDN. For different values of the coupling strengths, mMIR can be achieved for different numbers of added interconnections. In all cases studied, we also compute the global synchronization measure $\rho$ of the finally evolved BDN (for the details, see Materials and Methods, Subsection Synchronization Measures in BDNs) to allow for direct comparisons with mMIR and for the identification of relations between synchronization and information flow capacity in the BDNs.

\subsection*{Rescaling of Chemical and Electrical Couplings for Parameter Spaces of Networks with Different Eigenvalue Spectra}\label{subsection_a_rough_estimation_of_chemical_and_electrical_couplings_for_parameter_spaces}

Following Ref. \cite{Baptistaetal2010}, a rough estimation on the range of chemical and electrical couplings based on those used for the parameter space of another network that is capable of reproducing similar synchronous behaviors \cite{Baptistaetal2008A} and similar amounts of Kolmogorov-Sinai entropy \cite{baptista2011complex}, can be computed as following: Suppose we have produced a parameter space such as those of Fig. \ref{fig:zoomin_results_c.elegans_averaged_human_brain} showing behaviors of the BDN of the \textit{C.elegans} as a function of $g_n$ and $g_l$. Let us denote the maximum electrical coupling as $g_l^{\mathrm{C}}$, the maximum chemical as $g_n^{\mathrm{C}}$, the smallest positive eigenvalue of the Laplacian matrix of the electrical connections as $\omega_m^{\mathrm{C}}$ and the average degree of the chemical connections as $\bar{d}^{\mathrm{C}}$. Suppose now we want to compute a similar parameter space for another BDN. Let us denote by $g_l^{\mathrm{max}}$, $g_n^{\mathrm{max}}$, $\omega_m^{\mathrm{max}}$ and $\bar{d}^{\mathrm{max}}$ the corresponding values of the new parameter space. Then, for the new maximum couplings we have:
\begin{eqnarray}
g_n^{\mathrm{max}}&=&\bigg(\frac{\bar{d}^{\mathrm{C}}}{\bar{d}^{\mathrm{max}}}\bigg)g_n^{\mathrm{C}}\quad\mbox{(chemical coupling)}\label{g_n_max_new},\\
g_l^{\mathrm{max}}&=&\bigg(\frac{\omega_m^{\mathrm{C}}}{\omega_m^{\mathrm{max}}}\bigg)g_l^{\mathrm{C}}\quad\mbox{(electrical coupling)}\label{g_l_max_new}.
\end{eqnarray}
To arrive at Eqs. \eqref{g_n_max_new}, \eqref{g_l_max_new} using the results of Ref. \cite{Baptistaetal2010}, we have assumed that a network with an average degree $\bar{d}$ for its chemical connections behaves similarly to a network with the same degree for its chemical connections.

Equations \eqref{g_n_max_new} and \eqref{g_l_max_new} provide a rough estimation on the maximum coupling strengths that can be used for the new parameter spaces. Table S3 presents $\omega_m$, $\bar{d}$, $g_n^{\mathrm{max}}$ and $g_l^{\mathrm{max}}$ for the different BDNs considered in our work. Based on these rough predictions, we then identified as best matching ranges, those depicted in the figures of the paper. Based on the maximum values of the parameter space ranges of Fig. \ref{fig:zoomin_results_c.elegans_averaged_human_brain}(A) for the \textit{C.elegans} ($g_n^{\mathrm{C}}=0.3$ and $g_l^{\mathrm{C}}=2$), we get for the average humans $g_n^{\mathrm{max}}=0.21$ and $g_l^{\mathrm{max}}=1.71$, in good agreement with the maximum values of the ranges in Fig. \ref{fig:zoomin_results_c.elegans_averaged_human_brain}(C) and (D). For the model for brain network evolution we get $g_n^{\mathrm{max}}=1.14$ to 2.28 and $g_l^{\mathrm{max}}=0.97$ to 1.17 depending on the particular BDN, in accordance with the range of couplings used in Figs. \ref{fig:main_results_n60c6} and, 2 in Figure S2. Similarly, for the large version of our model for brain network evolution ($N_n=120$, $N_c=6$) we estimated $g_n^{\mathrm{max}}=2.16$ and 
$g_l^{\mathrm{max}}=2.9$, consistent with the maximum coupling strengths used in the last four panels of Fig. \ref{fig:spectral_distances}. Consequently, our methodology allowed us to identify regions of synchronization $\rho$ and upper bound for MIR, $I_c$, for the different BDNs of this work with similar functional and structural properties.

\subsection*{Analysis of Networks and Communities}\label{subsection_analysis_of_communities}

We initially identified the communities of the networks using the walktrap method \cite{Ponsetal2005} of the igraph software with six steps. The algorithm detects communities through a series of short random walks, with the idea that the vertices encountered on any given random walk are more likely to be within a community. The algorithm initially treats all nodes as communities of their own, then merges them into larger communities, and these into still larger, and so on. Essentially, it tries to find densely connected subgraphs (i.e. communities) in a graph via random walks. The idea is that short random walks tend to stay in the same community. Following this procedure we have been able to identify 6 communities in the \textit{C.elegans} BDN, 10 in human subject A$_1$, 5 in A$_2$, 9 in B, 6 in C, 10 in D and finally, 7 in E.

After this step, we computed various statistical quantities such as the global clustering coefficient, the average of local clustering coefficients, the mean shortest path, the degree pdf of the network and the small-worldness measure. The latter property is characterized by a relatively short minimum path length on average between all pairs of nodes in the network, together with a high clustering coefficient.

Even though small-worldness captures important aspects of complex networks at the local and global scale of the structure, it does not provide information about the intermediate scale. Properties of the intermediate scale can be more completely described by the community structure or modularity of the network \cite{Newman2004}. The modules of a complex network, also-called communities, are subsets of nodes that are densely connected to other nodes in the same module but sparsely connected to nodes belonging to other communities. Since nodes within the same module are densely intra-connected, the number of triangles in a modular network is larger than in a random graph of the same size and degree distribution, while the existence of a few links between nodes in different modules plays the role of topological shortcuts in the small-world topology. Systems characterized by this property tend to be small-world networks, with high clustering coefficient and short path length with respect to random networks.

To infer the small-worldness of a network or community, we first compute the mean local clustering coefficient $C$ of the network or community and the mean of the local clustering coefficients of one hundred randomly created networks $\langle C_r\rangle_{100}$ of the same degree pdf with the studied network and also, the mean shortest path of the studied network $L$ and the average of the mean shortest paths of the same one hundred random networks $\langle L_r\rangle_{100}$. Watts and Strogatz \cite{Wattsetal1998} measured that many real-world networks have an average shortest path length comparable to those of a random network ($L\sim L_r$), and at the same time a clustering coefficient significantly higher than expected by random chance ($C\gg C_r$). Then, they proposed a novel graph model, currently named the Watts-Strogatz model, with a small average shortest path length $L$, and a large clustering coefficient $C$. We adopt this as a working definition of a small-world network or community. Therefore, small-world networks are in between the limit cases of regular graphs with large $L$ and $C$ and random networks with small $L$ and $C$. To quantify small-worldness we use the ratios \cite{Basetetal2006}:
\begin{equation}
\mu=\frac{L}{\langle L_r\rangle_{100}},\quad \gamma=\frac{C}{\langle C_r\rangle_{100}},
\end{equation}
in such a way that, for a small-world network or community, we compute:
\begin{equation}
\sigma=\frac{\gamma}{\mu}>1,\label{sigma_value}
\end{equation}
being the small-worldness measure. The higher is $\sigma$ from unity for a given network or community, the better it displays the small-world property.

For completeness, we note that for the human subject D, the walk-trap community analysis with step equal to six detected eleven communities for which the last one comprised only one neuron. Hence, in all computations, we disregarded this community as a trivial case. We present the results of the above analysis in Table S4. We have performed all structural analyses of this paper using the igraph software. The evolution of the basic network of Figure S1, under the principle of the maximization of the information flow capacity, is able to capture behaviors of real brain connectivity networks such as those for the communities of the \textit{C.elegans} BDN, being their small-world structure a prominent reason. We found out that in all networks studied, the small-world measure $\sigma$ gets values much higher than unity (see Table S4), clearly indicating that they all display the small-world property, though in different degrees.

\subsection*{Brain Network Evolution Promotes Global no Hebbian-like and, Local Hebbian-like and no Hebbian-like Evolution Learning Processes}\label{subsection_Synchronization_vs_Information_Flow_Capacity_n60c6}

Here, based on the extraordinary functional similarities found so far, we focus on two characteristic behaviors of the model for brain network evolution defined in Materials and Methods, Subsection A Model for Brain Network Evolution Based on the Maximization of Information Flow Capacity, to reveal important modular behaviors and underlying learning evolution processes: We associate the first one to the case of high global synchronization and low information flow capacity (which we call case $\mathcal{A}$) and the other one to the opposite situation of low synchronization and high information flow capacity (case $\mathcal{B}$). For illustration purposes, we select from the proposed model for the brain network evolution of 60 neurons and six clusters of Materials and Methods, Subsection A Model for Brain Network Evolution Based on the Maximization of Information Flow Capacity, two finally evolved BDNs with the following coupling strengths: For case $\mathcal{A}$ the pair $g_n=0.2,\;g_l=1.8$ from the fifth realization (indicated by $\blacktriangle$ in Figs. \ref{fig:main_results_n60c6}, \ref{fig:spectral_distances}) and for case $\mathcal{B}$ the pair $g_n=0.9,\;g_l=1.5$ from the fourth realization (indicated by $\bullet$ in Figs. \ref{fig:main_results_n60c6}, \ref{fig:spectral_distances}). We have checked that the conclusions are valid independently of the realization and other similar pairs of coupling strengths. Also, our results reported here are independent on the initial small-world cluster configuration and initial conditions.

\begin{figure}[!ht]
\centering{
\includegraphics[scale=0.055]{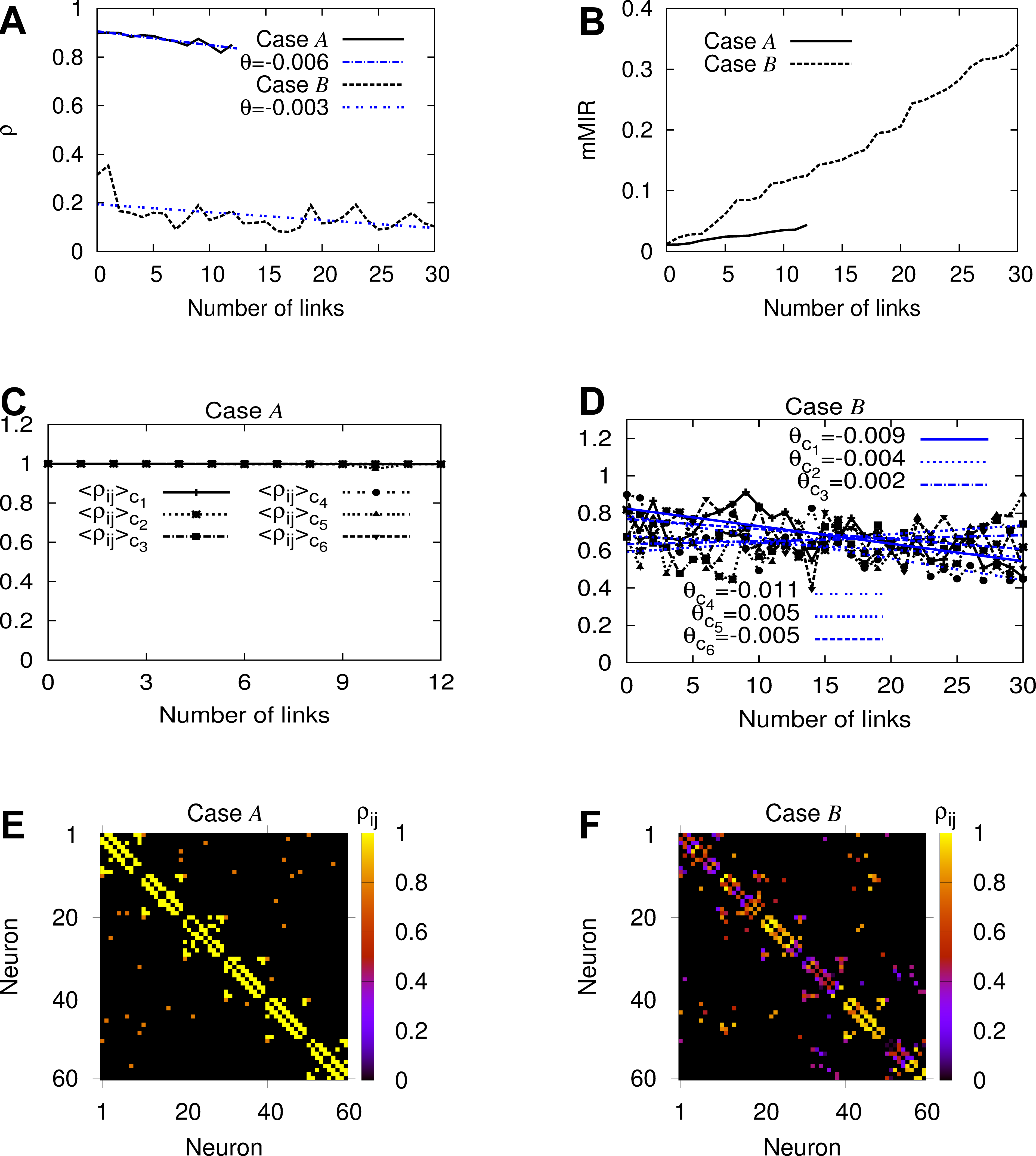}
}
\caption{\textbf{Brain network evolution promotes Hebbian-like and no Hebbian-like processes and, modular organization in the brain dynamical networks.} Panels (A), (B): Global synchronization $\rho$ and mMIR respectively for cases $\mathcal{A}$ and $\mathcal{B}$. $\theta$s are the slopes of the two dashed blue lines which are fitted to the black curves to demonstrate the decrease of the global neural synchrony during brain network evolution. Panels (C), (D) show how the average pair-wise synchronization $\langle\rho_{ij}\rangle_{c_l}$ of the clusters $c_l,\;l=1,\ldots,6$ changes during evolution, for cases $\mathcal{A}$, $\mathcal{B}$ respectively. Panels (E), (F) show the pair-wise neural synchronization level $\rho_{ij}$ of the finally evolved BDNs for the same cases. The horizontal axes (Number of links) correspond to the added links during brain network evolution that lead to the increase of the information flow capacity at each step. The caption of panel (D) for the black curves is the same as in panel (C). These results are for the studied model of brain network evolution with 60 neurons and six small-world clusters.}\label{fig:Synchronization_vs_Information_Flow_Capacity_n60c6}
\end{figure}

In panels (A), (B) of Fig. \ref{fig:Synchronization_vs_Information_Flow_Capacity_n60c6} we demonstrate the relation between global synchronization $\rho$ and mMIR for these cases. Case $\mathcal{B}$ of moderately low global synchronization (dashed black curve in panel (A)) and high information flow capacity (dashed black curve in panel (B)) is characterized by a larger number of added interconnections (i.e. 30) present in the finally evolved network with respect to case $\mathcal{A}$ of only 12 (corresponding to the solid black curves in panels (A), (B))! In both cases, $\rho$ of Eq. \eqref{z_t} for global synchronous behavior, has the tendency to decrease with different slopes (denoted by $\theta$ in Fig. \ref{fig:Synchronization_vs_Information_Flow_Capacity_n60c6}(A)) during brain network evolution as is evident by the fitting to the data in dashed blue lines. We attribute this behavior to an evolutionary brain network behavior, where global neural synchronization levels decrease during brain network evolution, reminiscent of a possible underlying global no Hebbian-like evolution process that promotes a decrease in global synchronization levels as new connections are added in the network.

We regard this as an important global property of the model for brain network evolution that needs to be further clarified, as we need to account for what happens on the local, cluster level as well. We thus use the following procedure to infer about the underlying learning rules for the clusters: During each step of the evolution process of the model of 60 neurons and six small-world clusters (which is a particular BDN), we compute $\langle\rho_{ij}\rangle_{c_l}$ based on Eq. \eqref{Cij} in Materials and Methods, to account for the average pair-wise synchronization of cluster $c_l,\;l=1,\ldots,6$. At the end of the time evolution, we have six such values for the clusters, lying in the interval $[0,1]$. We then record these values if the newly added chemical interconnection leads to an increase of the information flow capacity as depicted by the corresponding $I_c$ and disregard them if not. At the end of the process, we result with a relationship between $\langle\rho_{ij}\rangle_{c_l}$ and the added interconnections that maximizes $I_c$, for all clusters (see panels (C), (D) of Fig. \ref{fig:Synchronization_vs_Information_Flow_Capacity_n60c6}). Doing so, we can account for the underlying cluster learning processes. In particular, the terminology ``Hebbian-like'' is employed to represent a learning rule that not only involves quantities for synchronous events defined between pairs of neurons (see Eq. \eqref{Cij}), but also  a direct relation between synapse strength and synchronization increase. The ``no Hebbian-like'' terminology refers to a learning rule that involves not only a global measure of synchronous behavior (see Eq. \eqref{Cij}), but also refers to a phenomenon where the addition of synapses is accompanied by a decrease in the synchronization levels.

We present these results in Fig. \ref{fig:Synchronization_vs_Information_Flow_Capacity_n60c6}(C), (D). Following our considerations, we already know that globally, both finally evolved BDNs are following a no Hebbian-like learning rule. However, panels (C), (D) reveal a substantial difference for the synchronization behavior for pairs of neurons in the clusters in the two cases that allows us to assign different kinds of learning rules to them. For case $\mathcal {B}$ in panel (D), the slopes $\theta_{c_l}$ of the fitted lines (in blue) to the data $\langle\rho_{ij}\rangle_{c_l},\;l=1,\ldots,6$ (in black) show that the third and fifth cluster have the tendency to increase their internal synchronization level as the BDN evolves. In other words, neurons belonging to these two clusters follow a Hebbian-like learning process that comes in contrast to the no Hebbian-like learning behavior of neurons belonging to the other clusters as they show the opposite trend during brain network evolution, exhibiting negative slopes for the synchronization. These findings are in agreement with the conclusions drawn from panel (F) in which neurons belonging to the third and fifth cluster are seen to be more synchronized with respect to the others (in yellow and red), a situation that promotes the modular organization in the brain with different internal levels of synchronization. On the other hand, case $\mathcal{A}$ demonstrates a completely different situation in which the level of global synchronization decreases and cluster synchronization is very strong between neurons in all clusters during the whole brain network evolution, a behavior that results in a network whose neurons are unable to exchange but very small amounts of information. This is a case where the brain network can not transmit information by increasing its synaptic efficacy and thus, it is unable to learn new information!

Panels (E), (F) of Fig. \ref{fig:Synchronization_vs_Information_Flow_Capacity_n60c6} show the pair-wise synchronization level of the neurons denoted as $\rho_{ij}$ (see Eq. \eqref{Cij} in Materials and Methods, Subsection Synchronization Measures in BDNs) of the finally evolved BDNs for cases $\mathcal{A}$ and $\mathcal{B}$, resulting in networks of 12 and 30 chemical inter-links, respectively. Panel (F) of case $\mathcal{B}$ for moderately low global synchronization and high information flow capacity reveals a clusterized synchronization behavior. The third and fifth small-world clusters show much higher levels of internal synchrony (yellow and red points), i.e. synchronization of pairs of neurons in the same small-world community, with respect to the blue or dark blue points of low neural pair synchronization in other small-world clusters. The same also happens for a small number of pairs of neurons belonging to different clusters (off-diagonal points). One can notice that given the cluster that has the same internal level of synchronization, there can always be found two subnetworks, each belonging to one cluster, that have an equal amount of synchronization. This means that case $\mathcal{B}$ corresponds to a BDN that has clustered multilayer synchronization, where different clusters become functionally connected with different common behaviors. A situation reminiscent of findings in neuroscience that demonstrate the modular organization of the brain in which modular processors are sufficiently isolated and dynamically differentiated to achieve independent computations, but also globally connected to be integrated in coherent functions \cite{Meunieretal2001}. Our results for case $\mathcal{B}$ (see Fig. \ref{fig:Synchronization_vs_Information_Flow_Capacity_n60c6}(F)) come in contrast to those of $\mathcal{A}$ of high global synchronization and low information flow capacity shown in Fig. \ref{fig:Synchronization_vs_Information_Flow_Capacity_n60c6}(E) which demonstrates that all clusters and pairs of neurons attain almost the same state of almost complete synchronization, revealing a highly synchronized brain dynamical network that is not able however to exchange but only very small amounts of information between its different parts!

The no Hebbian-like mechanism is effectively similar to the unlearning anti-Hebbian mechanism of Crick and Mitchison \cite{Cricketal_1983} that proposes the elimination of unnecessary connections to prevent overload and to render the network more efficient. Both mechanisms lead however to more efficient networks, being in our case the evolved networks able to maximize their information flow capacity.

Typically, low synchronization implies $\lambda_1\approx\lambda_2$ leading to $I_c\approx0$. In critical points however, $I_c$ tends to be large as long as most of the oscillation modes are stable resulting in a situation where $\lambda_1>0$ and, $\lambda_2\approx0$ and positive. Thus, maximum $I_c$ at low synchronization corresponds to maximum $I_c$ near the critical point of the dynamics where $\lambda_1>0$ and, $\lambda_2\approx0$ and positive. Self-critical phenomena happen when the network has marginal Lyapunov exponents, in other words at the critical point that results in typically large $I_c$. In this context, our results are in agreement with the work in Ref. \cite{Yangetal_2012}, supporting our findings for the existence of Hebbian and no Hebbian-like learning mechanisms in the level of the communities (local synchronization) at self-criticality, which is responsible for the maximization of the information flow capacity of the evolved BDNs (low global synchronization), such as in case $\mathcal{B}$ (for a similar result see Ref. \cite{Turalskaetal2012}).

We do not study our BDNs under different stimuli. Our main hypothesis is that the final topology and behavior of an evolved BDN that maximizes MIR between its neurons is similar to real brain networks, such as those from the \textit{C.elegans} and human subjects. Since our results for the maximization of the information flow capacity of the evolved BDNs happen when self-critical phenomena emerge, they are in agreement with the results reported in Refs. \cite{Zare201380,Lukovic2014430}.

\subsection*{Structural Properties of the Model for Brain Network Evolution}\label{subsection_structural_properties_of_the_model_for_brain_evolution}

We study here the structural properties of the evolved BDNs and compare with those of the \textit{C.elegans} and human BDNs. As we have already demonstrated, they share common functional properties.

We present the results of this study in Fig. \ref{fig:statistical_analysis_brain_networks}. Particularly, panel (A) shows the degree probability distribution function (pdf($\bar{k}_i$)), panel (B) the clustering coefficient CC($\bar{k}_i$) as a function of the normalized degree $\bar{k}_i$, panel (C) the average degree $k_{nn}(\bar{k}_i)$ of the neighbors of nodes with degree $\bar{k}_i$ and panel (D) the network with its distinct communities depicted by different color-shaded neighborhoods, for case $\mathcal{A}$. In this context, $\bar{k}_i$ is the normalized with respect to the maximum, degree. Panels (E) to (H) show similar plots for case $\mathcal{B}$. Panels (I) to (L) are similar plots for the \textit{C.elegans} brain network and, panels (M) to (P) for human subject A$_1$. The plots in the second column show the correlation between different normalized degrees of the network whereas those of the third the tendency of the nodes of a certain normalized degree $\bar{k}$ to link with other nodes of a given degree. From the plots of the second column we observe that high degree nodes have the tendency to link with low degree nodes following an exponential dependence (with different exponents) implying disassortative mixing by degree. Our results from the second column of Fig. \ref{fig:statistical_analysis_brain_networks} suggest that evolving BDNs based on the maximization of the upper bound for MIR (cases $\mathcal{A}$ and $\mathcal{B}$) gives rise to disassortative mixing by degree meaning that high degree nodes are preferentially connected to other low degree nodes and low to high degree nodes.

It is worth noting that the structural properties of all human subjects are similar. Figures \ref{fig:statistical_analysis_brain_networks}(I) to (O) show that human subject A$_1$ has similar structural properties to the \textit{C.elegans} structure. Cases $\mathcal{A}$ and $\mathcal{B}$ seem also to present strong structural similarities, for the quantities considered in Fig. \ref{fig:statistical_analysis_brain_networks}, despite the profoundly different functional behaviors. We found out that case $\mathcal{B}$ of moderately low global synchronization and high information flow capacity is characterized by a big number of added interconnections (i.e. 30) present in the final network whereas case $\mathcal{A}$ by only 12! We have checked that this relationship between a large (small) number of added chemical inter-links with low (high) synchronization and large (low) information flow capacity happens for all similar cases of the parameter spaces. It is also noteworthy that in case $\mathcal{B}$ (see Fig. \ref{fig:statistical_analysis_brain_networks}(H)), the number of clusters of the finally evolved network is reduced by one (since we start with six and end up with five) and is in contrast with what is happening in case $\mathcal{A}$! Depending on the couplings which give rise to different functional behaviors, the brain network evolution model is capable of merging different communities, i.e. of restructuring the initial network configuration. Case $\mathcal{B}$ is a more globally connected network. 

The previous observation can be quantified in terms of the modularity of the network, i.e. by the strength of the division of the network into modules (groups, clusters or communities). Networks with high modularity have dense connections between the nodes within modules and sparse connections between nodes of different modules, being more clustered. We have used the igraph software for these computations. By applying this idea here, we find that the modularity of the final BDN of case $\mathcal{B}$ is 0.596, very close to the average modularity of the six human subjects $0.588\pm0.023$. In contrast, the modularity of case $\mathcal{A}$ is 0.702. For the sake of completeness, we also report the modularity of the \textit{C.elegans} topology which is 0.375, the smaller of all cases we considered. The last result shows a network with sparser connections between the nodes within the modules and denser between nodes of different modules! For the \textit{C.elegans}, the natural evolution process led to a smaller clusterization of its brain network.

The results of Fig. \ref{fig:statistical_analysis_brain_networks} suggest that the evolution process of a basic small-world clustered network is capable of generating evolved ones with similar structural properties to those for the \textit{C.elegans} and human BDNs. The structural similarity to the human brain is even more remarkable if the couplings of the network to be evolved are within the range that promotes high levels of information flow capacity such as in case $\mathcal{B}$.

\subsection*{Synchronization Measures in BDNs}\label{subsection_a_measure_of_synchronization_in_brain_networks}

Synchronous activity has been observed in neural systems and reported to be associated not only with pathological brain states \cite{Traub_1982} but also with various cognitive functions \cite{Mathias2011}. It has been found that burst synchronization of neural systems may be strongly influenced by many factors, such as coupling strengths and types \cite{Belykh_2005}, noise \cite{Buric_2007}, and the existence of clusters in neural networks \cite{Viana_2014}.

In this paper we use the order parameter $\rho$ to account for the synchronization level of the neural activity of the studied BDNs and of their communities \cite{Gardnesetal2010}. It is originated from the theory of measures of dynamical coherence of a population of $N_n$ oscillators of the Kuramoto type \cite{Kuramotoetal_2002} and, can be computed by a complex number $z(t)$ defined as:
\begin{equation}\label{z_t}
 z(t)=\rho(t)e^{\mathrm{i}\Phi(t)}=\sum_{j=1}^{N_n}e^{\mathrm{i} \phi_j(t)}.
\end{equation}
By taking the modulus $\rho(t)$ of $z(t)$, one can measure the phase coherence of the population of the $N_n$ neurons of the BDN, and by $\Phi(t)$ to measure the average phase of the population of oscillators. In this context, $\phi_i$ is the phase variable of the $i$-th neuron of the HR system \eqref{HR_model_Nneurons} given by its fourth equation. Actually, one averages over time $\rho(t)$ to obtain the order parameter $\rho=\langle\rho(t)\rangle_t$, the tendency of $\rho$ in time. A value of $\rho=1$ corresponds to complete synchronization of the oscillators, whereas $\rho=0$ to complete desynchronization.

We use Eq. \eqref{z_t}, adapted accordingly, wherever in the paper we need to compute the synchronization level of BDNs or clusters. In particular, in the case of BDNs, $N_n$ is the number of neurons of the BDN and $j$ runs through all $N_n$ neurons of that network whereas in the case of clusters, $N_n$ represents the number of neurons of the particular cluster and $j$ refers to the particular neurons which are members of this cluster.

We also compute and plot in Fig. \ref{fig:Synchronization_vs_Information_Flow_Capacity_n60c6}(C), (D), for cases $\mathcal{A}$ and $\mathcal{B}$ respectively of the model for brain network evolution of 60 neurons and six small-world clusters, the pair-wise neural synchronization by looking at the synchronization patterns between all pairs of neurons $i,j$ of the network as:
\begin{equation}\label{Cij}
 \rho_{ij}=\lim_{\Delta t\rightarrow \infty}\biggl |\frac{\boldsymbol{C}_{ij}}{\Delta t}\int_{\tau}^{\tau+\Delta t}e^{\mathrm{i}[\phi_i(t)-\phi_j(t)]}dt\biggr|,
\end{equation}
where $\boldsymbol{C}_{ij}$ is the adjacency matrix of the brain network and $\phi_i$ is the phase variable of the $i$-th neuron of system \eqref{HR_model_Nneurons}. $\rho_{ij}$ are bounded in the interval $[0,1]$, being $\rho_{ij}=1$ when neurons $i,j$ are fully synchronized and 0 when they are dynamically uncorrelated. To correctly compute $\rho_{ij}$, we take the averaging time large enough in order to obtain good measurements of the coherence degree of each pair. We similarly compute and plot in Fig. \ref{fig:Synchronization_vs_Information_Flow_Capacity_n60c6}(C), (D) for the same cases, the averaged quantity $\langle\rho_{ij}\rangle_{c_l},\;l=1,\ldots,6$ over the six clusters, where $i,j$ run through the neurons of each cluster.

\subsection*{Upper Bound for MIR}\label{subsection_Upper_Bound_for_MIR}

After Shannon's pioneering work \cite{Shannon1948} on information, it became clear that it is a very useful and important concept as it can measure the amount of uncertainty an observer has about a random event and thus provides a measure of how unpredictable it is. Another related concept to the Shannon entropy that can characterize random complex systems is the MI \cite{Shannon1948} which is a measure of how much uncertainty one has about a state variable after observing another state variable.

In Ref. \cite{Baptistaetal2012}, the authors have derived an upper bound for the MIR between two nodes or two groups of nodes of a complex dynamical network that depends on the two largest Lyapunov exponents $l_1$, $l_2$ of the subspace of the network formed by these nodes. In particular, they have shown that:
\begin{equation}\label{Ic_MIR}
 \mbox{MIR}\leq I_c=l_1-l_2,\;l_1\geq l_2,
\end{equation}
where $l_1$, $l_2$ are the two finite time and size Lyapunov exponents calculated in the bi-dimensional observation space of the two considered nodes \cite{Baptistaetal2012,Antonopoulosetal2014}, which typically should approach the two largest Lyapunov exponents $\lambda_1$, $\lambda_2$ of the dynamical network if it is connected and the time considered to calculate $l_1$, $l_2$ is sufficiently small. In our study, the upper bound $I_c$ for the MIR between any two nodes of the BDNs is effectively estimated by $I_c=\lambda_1-\lambda_2$ (i.e. $l_1=\lambda_1$ and $l_2=\lambda_2$) and will stand for the upper bound for the information transferred per time unit between any two nodes of the BDN (i.e. between the neurons), what represents the information flow capacity of the BDN. The phase spaces of the dynamical systems associated to the neural networks we study here are excessively highly multi-dimensional and thus, estimating an upper bound for the MIR using $\lambda_1$ and $\lambda_2$ calculated by the methods of Refs. \cite{Benettin1980a,Benettin1980b} instead of the MIR itself between all pairs of nodes, reduces enormously the computational complexity of the numerical calculations of this work. Besides, parameter changes that causes positive or negative changes in the MIR are reflected in the upper bound with the same proportion \cite{Baptistaetal2012}.

\subsection*{Normalized Laplacian Spectra}\label{subsection_Normalized_laplacian_spectra}

It is well-known that similarities between the structure of networks can be used for their classification \cite{Wilson_2008}. The architecture of brain networks that describe the organization of maps of connections between neurons and brain elements at a systems level can be achieved by examining the eigenvalue spectrum of the normalized Laplacian of the connectome \cite{Chung1996,Lange_2014}. In our study, the connectome is given by the adjacency matrix of the brain network and thus we compute the eigenvalues of the normalized Laplacian based on this matrix. The eigenvalues $\nu_{i},i=1,\ldots,N_n$ of the normalized Laplacian matrix $\boldsymbol{L}$ are in $[0,2]$ which helps to compare networks of different sizes, and is defined as:
\begin{equation}
\boldsymbol{L}_{ij} =
\begin{cases}
1 & \quad \text{if $i$ = $j$},\\
-\frac{1}{k_i} & \quad \mbox{if $i,j$ are connected},\\
0 & \quad \text{otherwise},\\
\end{cases}
\end{equation}
where $i,j$ represent any two nodes of the brain network, $\boldsymbol{L}_{ij}$ the link between nodes $i,j$ and $k_i$ the degree of node $i$. Therefore, the Laplacian spectrum of the network is given by the set of all eigenvalues of $\boldsymbol{L}$, namely by its eigenspectrum.

\paragraph*{Spectral Plot}

The spectral plots were obtained from a smoothed eigenvalue distribution $\Gamma(x)$ that consists of eigenvalue frequencies convolved with a Gaussian kernel \cite{Wilson_2008}:
\begin{equation}
\Gamma(x) = \sum\limits_{i=1}^{N_n} \frac{1}{\sqrt{2\pi\sigma^{2}}}\exp\bigg(-\frac{\left|x - \nu_{i}^{2}\right|}{2\sigma^{2}} \bigg),\label{Gamma_function}
\end{equation}
where $N_n$ is the number of eigenvalues of $\boldsymbol{L}$ and $\sigma$ a smoothing function. In our study, we used $\sigma=0.015$ to smooth out appropriately the spectral plots. For these plots, a discrete smoothed spectrum was used in which $\Gamma$ had steps of 0.001 and the distribution was normalized in such a way that the total eigenvalue frequency is unity.

\paragraph*{Spectral Graph Distance}\label{spectral_distance}

The similarity distance between spectral plots was quantified using a spectral distance measure, based on the distance measure introduced in Ref. \cite{Wilson_2008}, and is defined as the average Euclidean distance between two spectral plots $\Gamma_{1}$ and $\Gamma_{2}$:
\begin{eqnarray}
D(\Gamma_{1},\Gamma_{2})&=& \frac{1}{k+1}\sum\limits_{i=0}^k \min_{j} \bigg(\sqrt{[\Gamma_{1}(i)- \Gamma_{2}(j)]^{2}+[i-j]^{2}} \bigg) \nonumber\\ &+& \frac{1}{k+1}\sum\limits_{j=0}^k \min_{i} \bigg(\sqrt{[\Gamma_{1}(i)- \Gamma_{2}(j)]^{2}+[i-j]^{2}} \bigg),\label{similarity_distance}
\end{eqnarray}
where $\Gamma (i)$ is the discrete, normalized and smoothed eigenvalue distribution of Eq. \eqref{Gamma_function} and the number of intervals $k$ was set to 2000. The distance function \eqref{similarity_distance} depends on the scaling of the axes, meaning that different scales result in different distances. Therefore, it is not an invariant distance measure between two networks but only serves as a tool to underpin the visual results in a quantitative manner.

To generalize the important relation between function and structure in BDNs, we plot in panels (C) to (F) in Fig. \ref{fig:spectral_distances} the results of a similar study for one realization of a double-sized model for brain network evolution based on 120 HR coupled neurons arranged in six small-world clusters. Panels (C) and (D) show the parameter spaces for the global synchronization $\rho$ and information flow capacity mMIR of the big BDN which reproduce well the behavior of the small version of the model for brain network evolution of Fig. \ref{fig:main_results_n60c6}(A), (B) and allow for similar functional conclusions to be drawn. The relation between structural and functional properties for the double-sized model can be seen in the last two panels where we present the parameter spaces for the spectral distances when compared with the \textit{C.elegans} (panel (E)) and averaged humans (panel (F)). Again, couplings that promote low global neural synchronization and high information flow capacity give rise to the biggest possible spectral similarity of the double-sized brain network evolution model with the \textit{C.elegans} and human brain networks, the same conclusion drawn for the small version of the same model! We therefore verify our hypothesis that finally evolved BDNs with neurons that can potentially exchange the highest levels of information are the BDNs with topologies closer to the brain networks of the \textit{C.elegans} and humans.

\section*{Discussion}

In this paper we propose a working hypothesis, and provide evidence, that neural networks that evolve based on the principle of the maximization of their internal information flow capabilities produce networks whose functional behavior and topology are similar to those features observed in dynamical neural networks whose topology is provided by the \textit{C.elegans} and humans.

Our hypothesis goes along the lines of the infomax theory that proposes that the brain evolves by maximizing the mutual information between external stimuli and its response. When maximizing the internal information flow capacity, we are creating a network capable of processing information about external stimuli for which its information content is smaller than the information flow capacity of the evolved network. Notably, the brain evolves by the action of input signals. Our working hypothesis simplifies enormously the complexity of the involved calculations and, allow us to understand function and behavior in the brain, without the need to externally perturb non-autonomous neural networks.

We have been able to show that our evolved brain networks present similar synchronization and information flow capacity behaviors with the ones found for the simulated dynamical networks for the structure of the \textit{C.elegans} and human brain. Moreover, we have shown that BDNs evolved with coupling strengths that maximize the information flow capacity are the ones that have the smallest spectral graph distance from the BDNs of the \textit{C.elegans} and humans, and that, during the growing process, their MIR increase is related to moderately low amounts of global neural synchronization. Actually, the global neural synchronization levels decrease during brain network evolution, revealing an underlying global no Hebbian-like evolution process (where synapse strength leads to global decay of synchronization) driven by a mix of local no Hebbian-like learning rules for neurons in some clusters and by Hebbian-like learning rules in neurons belonging to other clusters where synapse strength leads to cluster synchronization. In this context, the no Hebbian-like mechanism is effectively similar to the unlearning anti-Hebbian mechanism as both lead to more efficient networks, in the sense that in our case the evolved networks are able to maximize their information flow capacity.

We note that if other models than Hindmarsh-Rose will be used, such as the Morris-Lecar, Izhikevich, or spiking map-based neural models, then the parameter regions that maximize information flow capacity and minimize synchronization (or vice versa) will be different, however we expect that this would not change the main results and conclusions of this work in the sense that brain network evolution based on the maximization of information flow capacity will lead to similar topologies, behaviors and relations for the evolved networks. For the human subjects, the graphs represent functionally connected brain regions. The Wilson-Cowan model could be appropriate to model the human brain, whereas it will not be suitable to model the \textit{C.elegans} brain. As we have done here, using neurons to represent nodes in the human connectome do not reproduce the real dynamics of the brain but gives us a mean to compare results with the \textit{C.elegans} and the evolved networks.

In relation to our work, the maximization of the information flow capacity for low synchronization corresponds to the critical point of the dynamics in which self-critical phenomena occur when the second or larger Lyapunov exponents of the BDNs are marginally positive.

Finally, our results support further the hypothesis made in Ref. \cite{Yamaguti2014} that maximization of the information flow capacity can serve as a principle for the development of heterogeneous structures in brain dynamical networks, such as the neocortex of mammalian brains.

\section*{Acknowledgments}
This work was performed using both the Maxwell high performance computing cluster and the ICSMB cluster of the University of Aberdeen. We are in debt with enlightening discussions with Dr. Iris Oren.

\section*{Author Contributions}
Conceived and designed the experiments: CGA MSB. Performed the experiments: CGA. Analyzed the data: CGA SS MSB. Contributed reagents/materials/analysis tools: CGA SS SESP MSB.
Wrote the paper: CGA SS MSB.


%
%
%

\providecommand{\noopsort}[1]{}\providecommand{\singleletter}[1]{#1}%

\pagebreak

\setcounter{equation}{0}
\setcounter{figure}{0}
\setcounter{table}{2}
\setcounter{page}{1}
\makeatletter
\renewcommand{\theequation}{S\arabic{equation}}
\renewcommand{\thefigure}{S\arabic{figure}}
\renewcommand{\thetable}{S\arabic{table}}

\begin{center}
\textbf{Do Brain Networks Evolve by Maximizing their Information Flow Capacity?}
\vspace{1cm}\\
Chris G. Antonopoulos$^{1}$, Shambhavi Srivastava$^{1}$, Sandro E. de S. Pinto$^{2}$ and\\Murilo S. Baptista$^{1}$
\vspace{1cm}\\
$^{1}$Department of Physics (ICSMB), University of Aberdeen, SUPA, Aberdeen, United Kingdom
\\
$^{2}$Departamento de F\'isica, Universidade Estadual de Ponta Grossa, Paran\'a, Brazil
\par\end{center}

\section*{Supporting Information}


\subsection*{S1}\label{S1_Fig}
\begin{figure}[!ht]
\centering{
\includegraphics[scale=0.08]{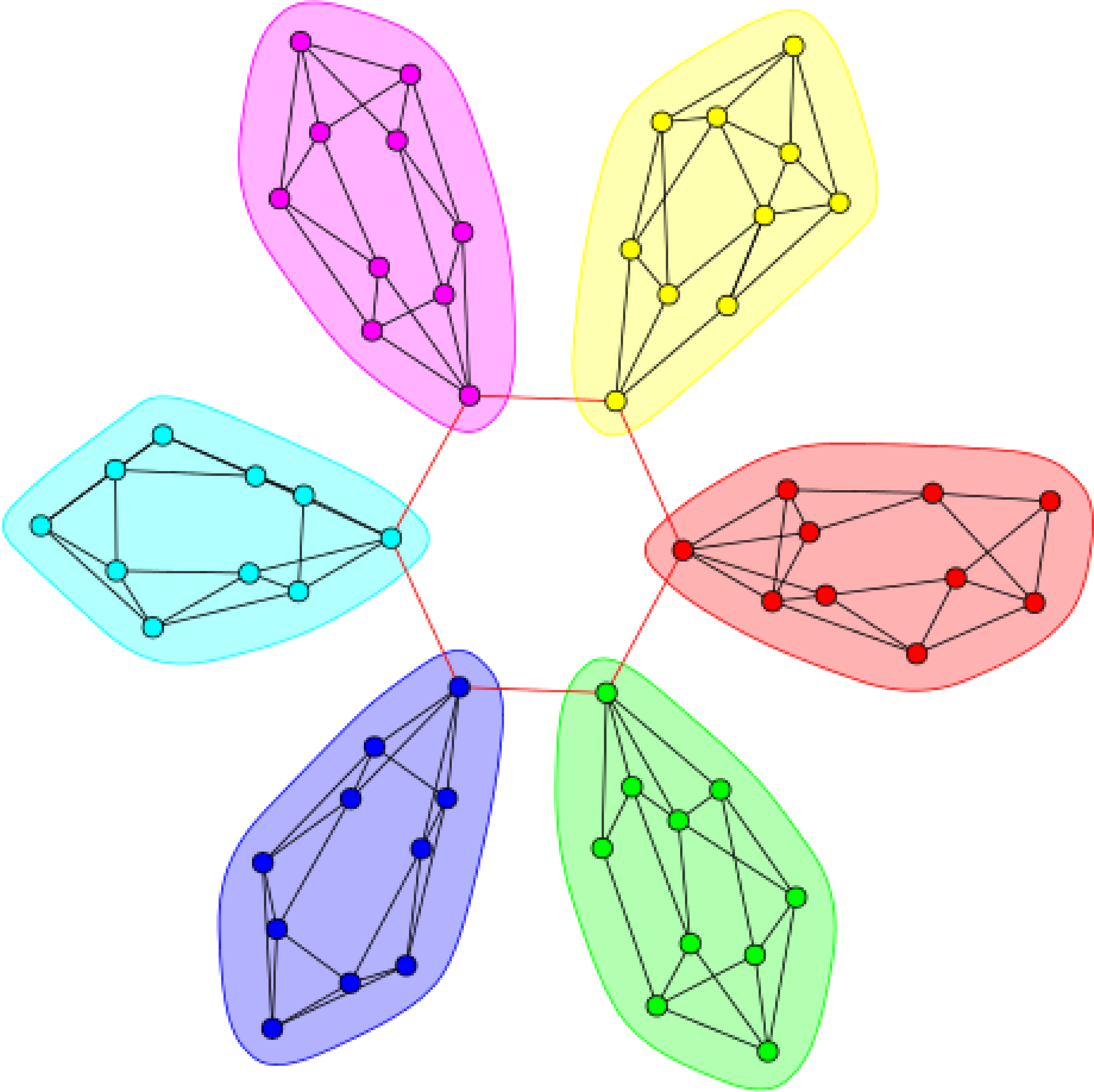}
}
\caption{\textbf{An example (first of the five realizations) of a starting small-world network topology for the proposed brain network evolution model.} It comprises $N_n=60$ neurons arranged in $N_c=6$ small-world clusters. The red ring consists of chemical excitatory connections that link all clusters of the network. Within each cluster, depicted by a differently color-shaded area of intra-connected neurons, we consider solely electrical connections denoted by black edges.}
\end{figure}

\newpage
\subsection*{S2}\label{S2_Fig}
\begin{figure}[!ht]
\centering{
\includegraphics[scale=0.046]{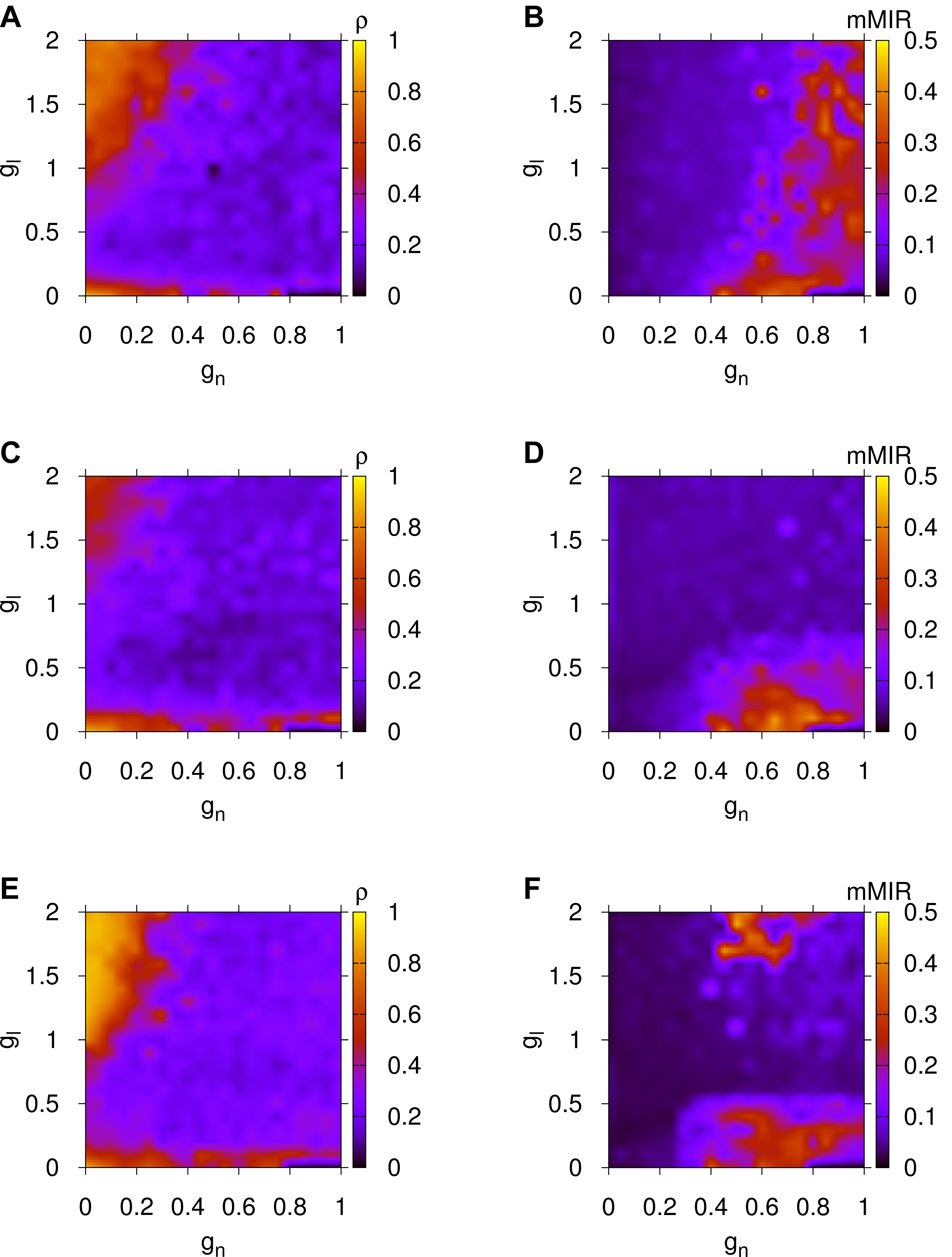}
}
\caption{\textbf{Global synchronization and information flow capacity properties for the evolved BDNs for different initial community structures.} (A): Parameter space for the synchronization $\rho$ and (B): for mMIR for six totally random (Erd\H{o}s-R\'enyi) clusters. (C) and (D) are similar plots but for six scale-free (Barab\'asi-Albert) clusters. (E) and (F) are for six star clusters perturbed by 20\%. All plots are from the model for brain network evolution of 60 neurons and six clusters. Here, $g_n$ is the chemical and $g_l$ the electrical coupling of Eqs. (2) of the paper.}
\end{figure}

\newpage
\subsection*{S3}\label{S1_Table}
\begin{table}[ht]
\caption{\textbf{Smallest positive eigenvalue $\boldsymbol{\omega_m}$ of the Laplacian matrix $\boldsymbol{G}$ of the electrical connections and average degree of chemical connections $\boldsymbol{\bar{d}}$ of the adjacency matrix $\boldsymbol{B}$ for the BDNs considered in this work.} These values were used to provide rough estimates for the extend of the couplings of the parameter spaces, $g_n^{\mathrm{max}}$ and $g_l^{\mathrm{max}}$, based on those of the zoomed-in parameter space of the \textit{C.elegans} of Fig. 2 of the main manuscript (first row of the Table).}
\centering
\begin{tabular}{| c | c | c | c | c |}
 \hline
\rule{0pt}{12pt} Brain network& $\omega_m$ & $\bar{d}$ & $g_n^{\mathrm{max}}$ & 
$g_l^{\mathrm{max}}$ \\ \hline
\textit{C.elegans} & 0.66 & 2.28 & 0.3 & 2 \\
\hline
Case $\mathcal{A}$ ($N_n=60$, $N_c=6$)& 1.35 & 0.3 & 2.28 & 0.97 \\
\hline
Case $\mathcal{B}$ ($N_n=60$, $N_c=6$)& 1.12 & 0.6 & 1.14 & 1.17 \\
\hline
$N_n=120$, $N_c=6$ & 0.45 & 0.32 & 2.16 & 2.9 \\
\hline
Average over the six human subjects & 0.8$\pm$0.15& 3.28$\pm0.55$ & 0.21$\pm0.038$ & 1.71$\pm0.37$ \\
\hline
\end{tabular}
\end{table}

\subsection*{S4}\label{S2_Table}
\begin{table}[ht]
\caption{\textbf{Small-worldness measure $\boldsymbol{\sigma}$ for the \textit{C.elegans} and human brain networks and for their communities.}}
\centering
\begin{tabular}{| c | c |*{6}{c|}}
 \hline
  & \textit{C.elegans} & \multicolumn{6}{c|}{Human subject} \\ \hline
  &   & A$_1$&A$_2$&B&C&D&E  \\
\hline
Full network          & 2.87 & 8.22 & 7.50 & 8.00 & 7.92 & 7.89 & 7.50\\
\hline
Community 1  & 2.06 & 2.92 & 3.79 & 2.75 & 3.52 & 3.09 & 3.74\\
\hline
Community 2  & 1.88 & 2.95 & 3.68 & 3.53 & 2.95 & 2.94 & 2.95\\
 \hline
Community 3  & 2.54 & 2.93 & 3.35 & 3.06 & 2.76 & 3.81 & 3.13\\
\hline
Community 4  & 2.29 & 3.09 & 2.99 & 3.19 & 3.56 & 3.07 & 3.58\\
 \hline
Community 5  & 2.37 & 2.97 & 2.92 & 2.89 & 2.85 & 2.82 & 2.91\\
\hline
Community 6  & 2.50 & 2.77 &         & 2.67 & 2.86 & 3.03 & 3.00\\
\hline
Community 7  &         & 2.83 &         & 2.46 &         & 2.86 & 2.94\\
\hline
Community 8  &         & 2.87 &         &  2.82 &         & 3.09 &       \\
\hline
Community 9  &         & 3.09 &         & 2.69 &         & 3.09 &       \\
\hline
Community 10 &         & 2.78 &        &         &         & 2.83 &       \\
\hline
\end{tabular}
\end{table}

\end{document}